\documentclass[twocolumn]{aastex63}

\usepackage{rotating}
\usepackage{amsmath}
\usepackage{soul}
\DeclareMathOperator{\sech}{sech}

\newcommand{\tctff}{t_{\rm cool}/t_{\rm ff}}
\newcommand{\Msun}{\mathrm{M_\odot}}

\graphicspath{{./}{figures/}}

\begin{document}

\title{Seeking Self-Regulating Simulations of Idealized Milky Way-Like Galaxies}

\correspondingauthor{Claire Kopenhafer}
\email{kopenhaf@msu.edu}

\author[0000-0001-5158-1966]{Claire Kopenhafer}
\affiliation{Department of Physics \& Astronomy, Michigan State University}
\affiliation{Department of Computational Mathematics, Science, \& Engineering, Michigan State University}
\author[0000-0002-2786-0348]{Brian W. O'Shea}
\affiliation{Department of Physics \& Astronomy, Michigan State University}
\affiliation{Department of Computational Mathematics, Science, \& Engineering, Michigan State University}
\affiliation{Facility for Rare Isotope Beams, Michigan State University}
\author[0000-0002-3514-0383]{G. Mark Voit}
\affiliation{Department of Physics \& Astronomy, Michigan State University}

\begin{abstract}
    Precipitation is potentially a mechanism through which the circumgalactic medium (CGM) can regulate a galaxy's star formation. 
    Here we present idealized simulations of isolated Milky Way-like galaxies intended to examine the ability of galaxies to self-regulate their star formation, particularly via precipitation. Our simulations are the first CGM-focused idealized models to include stellar feedback due to the explicit formation of stars.
    We also examine the impact of rotation in the CGM. Using six simulations, we explore variations in the initial CGM $\tctff$ ratio and rotation profile. 
    Those variations affect the amount of star formation and gas accretion within the galactic disk.
    Our simulations are sensitive to their initial conditions, requiring us to gradually increase the efficiency of stellar feedback to avoid destroying the CGM before its gas can be accreted. Despite this
    gradual increase, the resulting outflows still evacuate large, hot cavities within the CGM and even beyond $r_{200}$. Some of the CGM gas avoids interacting with the cavities and is able to feed the disk along its midplane, but the cooling of feedback-heated gas far from the midplane is too slow to supply the disk with additional gas.
    Our simulations illustrate the importance of physical mechanisms in the outer CGM and IGM for star formation regulation in Milky Way-scale halos.
\end{abstract}


\section{Introduction} \label{sec:intro}

Galaxies obey a number of scaling relations that suggest they have a mechanism for regulating their star formation rates (SFRs). These scaling relations include the star-forming main sequence \citep[e.g.,][]{popesso22-MainSequenceStarforming, sherman21-ShapeScatterGalaxy,renzini15-ObjectiveDefinitionMain}  and the stellar mass--halo mass (SMHM) relation \citep{somerville15-PhysicalModelsGalaxy, behroozi19-UNIVERSEMACHINECorrelationGalaxy}. Gas must be continually supplied to a star-forming galaxy as its own interstellar medium (ISM) only contains enough gas to fuel star formation for a few gigayears. The circumgalactic medium (CGM) has gained much attention as a possible locale for self-regulation as it mediates both inflowing and outflowing gas in addition to functioning as a substantial baryonic reservoir in its own right \citep{tumlinson17-CircumgalacticMedium}.

One potential self-regulation mechanism is known as ``precipitation'' \citep{voit15-PrecipitationRegulatedStarFormation}. Under this framework, feedback (either stellar or from an AGN) works to maintain the median $\tctff$ ratio around 10. Throughout this paper we'll refer to the median value $\tau$ of $\tctff$ in the CGM as the ``precipitation limit parameter.'' Importantly, $\tau$ is the median of a $\tctff$ \textit{distribution} for the CGM gas that may have a broad dispersion \citep{voit21-GraphicalInterpretationCircumgalactic}. When $\tau \sim 10$, the $\tctff$ distribution allows for the formation of a multiphase medium. If $\tau$ declines, additional cold dense gas is able to accrete onto the galaxy, driving a burst of star formation and feedback. This feedback in turn heats and inflates the CGM, lowering its density and reducing its ability to cool. Conversely, a CGM driven above $\tau \sim 10$ should feed less cold, dense gas into the galaxy. This reduction of the cold gas supply then stalls feedback until the CGM is again able cool and contract, thereby lowering $\tau$.

Precipitation as a regulation mechanism was originally proposed for galaxy cluster cores heated by active galactic nuclei (AGN). It was motivated by observations suggesting that that the black hole fueling rate depends on the development of a multiphase medium \citep{pizzolato05-NatureFeedbackHeating,cavagnolo08-EntropyThresholdStrong,voit08-ConductionStarFormation}, which happens in simulations when a CGM in approximate thermal balance becomes thermally unstable \citep{pizzolato10-SolvingAngularMomentum,mccourt12-ThermalInstabilityGravitationally,sharma12-ThermalInstabilityFeedback}. In simulated multiphase media, cold clouds are able to precipitate out of the hot, ambient medium when  $t_{\rm cool}  \, \lesssim \, 10 \, t_{\rm ff}$ \citep{gaspari12-CauseEffectFeedback,gaspari13-ChaoticColdAccretion,gaspari15-ChaoticColdAccretion,li14-ModelingActiveGalactic,li14-ModelingActiveGalactica,prasad15-CoolCoreCycles}. If the medium is sufficiently turbulent, these cold clouds are then able to accrete chaotically toward the center, and in the systems originally considered, they trigger a strong AGN feedback response. This feedback then heats and expands the CGM, raising $t_{\rm cool}$ and diminishing accretion. See \citet{donahue22-BaryonCyclesBiggest} for an in-depth review of this process.

Regulation via precipitation is observationally supported in both clusters \citep{voit15-CoolingTimeFreefall} and elliptical galaxies \citep{frisbie20-PropertiesHotAmbient}. The physical principles that underpin precipitation regulation are agnostic to the source of feedback in a galaxy, and so could extend to smaller galaxies not dominated by AGN feedback. This possibility is supported by recent theoretical \citep{voit19-CircumgalacticPressureProfiles} and observational \citep{babyk18-UniversalEntropyProfile} developments. 

In this work, we explore the viability of precipitation as a regulatory mechanism for star formation in Milky Way-like galaxies lacking AGN activity. Self-regulation is broadly defined as a balance between gas accretion on the one hand and stellar feedback on the other \citep{schaye10-PhysicsDrivingCosmic,bouche10-ImpactColdGas, dave12-AnalyticModelEvolution,zaragoza-cardiel19-DetectionSelfregulationStar}. This balance might ultimately be what places galaxies on the stellar mass-halo mass relation \citep{voit15-PrecipitationRegulatedStarFormation, mitchell21-HowGasFlows}.

Spiral galaxies like the Milky Way tend to live in complex environments where they are continuously interacting with companion galaxies and cosmological filaments. Such environments make it difficult to isolate the interactions of the CGM and its host galaxy and to understand the long-term behavior of an undisturbed Milky Way-like halo. We therefore use idealized simulations of an isolated galaxy for our investigation, seeking to understand what a Milky Way-like galaxy would look like without external interference. Idealized simulations also allow us to maintain a constant dark matter halo mass, as changing the dark matter mass should eventually change the stellar mass of the galaxy according to the SMHM relation. With idealized simulations, we are also able to achieve higher spatial resolution in the CGM for reasonable computational cost. High CGM resolution broadens the distributions of gas densities and temperatures \citep{corlies20-FiguringOutGas} and increases the overall amount of neutral hydrogen \citep{vandevoort19-CosmologicalSimulationsCircumgalactic}. Both of these effects should have an impact on precipitation \citep{voit21-GraphicalInterpretationCircumgalactic}.

Historically, idealized simulations of isolated Milky Way-like galaxies have fallen into two types: simulations of a star-forming gaseous disk with no CGM, and simulations of a CGM with no explicit star formation. The former include the AGORA simulations of \citet{kim16-AGORAHighresolutionGalaxy}, works using the AGORA initial conditions such as \citet{butsky18-RoleCosmicrayTransport} and \citet{shin21-HowMetalsAre}, and others \citep{benincasa16-AnatomyStarformingGalaxy}. Instead of a CGM, the gas disk of these simulations is surrounded by a very low density medium with very long cooling times and low total mass. In the latter category are the simulations from \citet{fielding17-ImpactStarFormation} and \citet{li20-HowSupernovaeImpact}. These CGM-focused works include stellar feedback, but this feedback is tied to the gas flow rate through an inner boundary rather than the explicit formation of stars in a gas disk. This work is the first to combine both approaches, modeling star formation and the resulting feedback in the context of both a gas disk and its surrounding non-uniform CGM to create a fully self-consistent picture of how the galaxy and its CGM interact. This is particularly important for modeling the inner CGM and the disk-halo interface.

We detail our simulation setup in Section \ref{sec:method}. Then, in Section \ref{sec:disk}, we examine how the galactic disk is affected by our simulation setup and its variations. In Section \ref{sec:inflow-outflow}, we follow the movement of gas between the disk and the CGM by tracing inflows, outflows, and the CGM's gas supply. Section \ref{sec:cgm} looks at the structure of the CGM in our simulations. We discuss various aspects of our simulations in Section \ref{sec:discussion}. Finally, we summarize and conclude our work in Section \ref{sec:conclusions}.

\section{Simulation Setup} \label{sec:method}

We perform idealized simulations of isolated, Milky Way-like galaxies and their circumgalactic media. Our simulations are performed with the Eulerian astrophysical hydrodynamics code Enzo \citep{bryan14-EnzoAdaptiveMesh, brummel-smith19-ENZOAdaptiveMesh}. Enzo models stellar populations with particles and we include feedback from Type II supernova, as will be discussed more in Section \ref{sec:SF-FB}.

We employ a 9-species, non-equilibrium primordial chemistry network through the Grackle cooling and chemistry library \citep{smith17-GRACKLEChemistryCooling}. This network includes all ionization states of H, He, and $\rm H_2$, including $\rm H^-$, as well as the density of free electrons. Grackle provides cooling rates for H, He, and metals through precomputed Cloudy\footnote{A slightly modified version of Cloudy 10 is used which saves outputs with more precision.} tables \citep{ferland13-2013ReleaseCloudy}. These tables include the extragalactic UV background of \citet{haardt12-RadiativeTransferClumpy}, which is fixed at $z=0$ for the purposes of our simulations.

We adopt a static NFW dark matter potential \citep{navarro96-StructureColdDark} with $M_\mathrm{200} = 10^{12}\ \Msun$ and a concentration of 10. We define the edge of the dark matter halo using $r_{200}\approx206$~kpc, which is the radius enclosing a density that is 200 times the critical density. 

On top of this we also impose a static stellar disk potential from \citet{miyamoto75-ThreedimensionalModelsDistribution}. For this potential we adopt $M_\ast = 5.8\times10^{10}\ \Msun$, which is the uncertainty-weighted average of observations compiled in Table 6 of \citet{cote16-UncertaintiesGalacticChemical}. We also set the radial and vertical scale heights of the stellar potential to match those of the gas for simplicity (see Table \ref{tab:ICs}). This potential is included because, though the simulation is not initialized with any star particles, we want to model the gravitational effects that the pre-existing stellar population of a $z\sim0$ Milky Way-like galaxy would provide. In particular, this potential influences the circular velocity of the disk gas. Star particles are allowed to form as the simulation evolves, effectively allowing the stellar component of the disk to grow from $M_\ast = 5.8\times10^{10}\ \Msun$.

Our galaxy and its background potentials are fixed to the center of a ({1,638.4}~kpc)$^3$ box with periodic boundary conditions. While periodic boundary conditions are not physically motivated, they are an easy way to ensure the total mass of the box is conserved. Additionally, the width of the box is about 8 times $r_{200}\approx206$~kpc, ensuring that the periodic conditions will not lead to any unexpected consequences.

The initial construction of the disk and CGM are covered in Sections \ref{sec:disk-ICs} and \ref{sec:CGM-ICs}, respectively. Section \ref{sec:SF-FB} covers the details of our star formation and feedback models. In Section \ref{sec:amr} we discuss our simulations' approach to AMR. Finally, Section \ref{sec:variants} covers the multiple simulation variants we run.

\subsection{Initial Conditions} \label{sec:ICs}

\begin{figure}
    \centering
    \includegraphics[width=\linewidth]{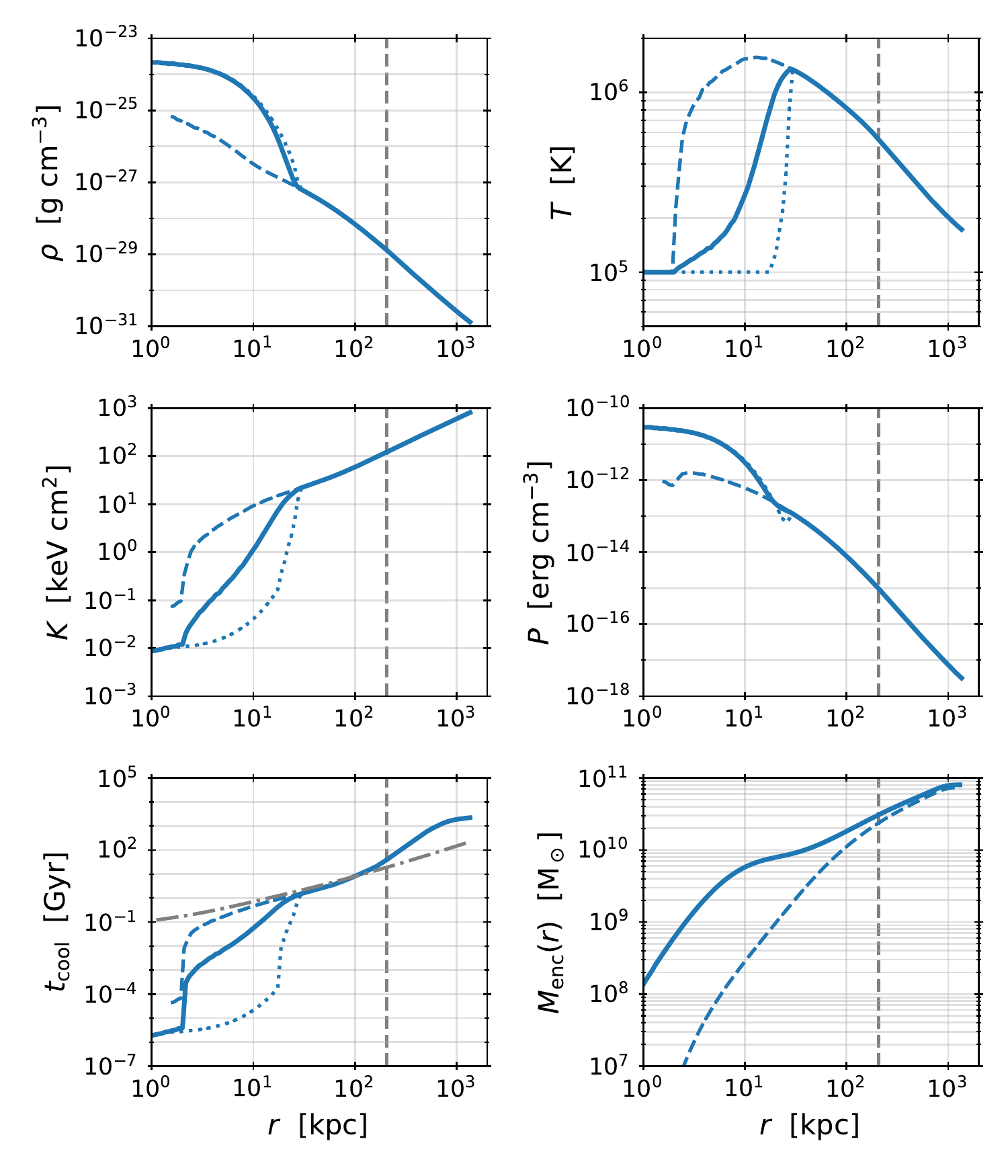}
    \caption{In rows from top left, as a function of radius: initial mass-weighted average density, temperature, entropy, pressure, and cooling time, and total enclosed mass. The independent contributions of the disk (dotted) and inner CGM (dashed) are shown for small radii, except for the enclosed mass. For the enclosed mass, the entire CGM is shown as a dashed line. Though the smallest cell size in the simulation is 100~pc, the x-axis does not extend this far because of the low number of cells. The vertical dashed line marks $r_{200} \approx 206$~kpc. The gray dot-dashed line shows where $\tctff=10$.
    }
    \label{fig:ics}
\end{figure}

The bulk values used to initialize our simulation can be found in Table \ref{tab:ICs}. Initial mass-averaged profiles are shown in Figure \ref{fig:ics}. The profiles are split into disk and CGM components at small radii.

Our stellar mass is taken as the uncertainty-weighted average of the observations in Table 6 of \citet[]{cote16-UncertaintiesGalacticChemical}. We then use the findings of \citet{peeples14-BudgetAccountingMetals} to derive the ISM mass and metallicity based on this stellar mass.  The disk metallicity is $Z_{\rm ISM}=2.3 \mathrm{\ Z_\odot}$, assuming $M_\mathrm{dust} = M_{Z,\mathrm{dust}}$. \citet[]{peeples14-BudgetAccountingMetals} suggests $M_{\rm ISM} = 9.8 \times 10^9\ \Msun$, but we lower this to $ 7 \times 10^9\ \Msun$ to help minimize the initial burst of star formation experienced by the simulation; see Section \ref{sec:FB-ramp} for more discussion.

The CGM is initialized to a constant uniform metallicity of $0.3\ \mathrm{Z_\odot}$. This is the median value from \citet{prochaska17-COSHalosSurveyMetallicities}. The metallicity of the CGM changes, however, as stellar populations inject supernova feedback and drive outflows (see Section \ref{sec:SF-FB}).

\begin{table}[t]
\centering
\begin{tabular}{llr}
\toprule
Quantity & Name & Value\\
\tableline
$M_{200}$ & virial mass & $1.0\times 10^{12}\ \Msun$\\
$C_{\rm NFW}$ & dark matter concetration & 10\\
$r_{200} $ & virial radius & 206~kpc\\
$M_\ast$ & background stellar mass & $5.8\times 10^{10}\ \Msun$\\
$M^\prime_{ISM}$ & effective\footnote{The actual initial disk mass ends up being slightly higher than the value of the parameter used in Equation \ref{eq:disk}, which is $M_{\rm ISM}=5.0\times10^9$. The actual disk mass that results is determined from a temperature cut, as the initial disk is isothermal and distinctly cooler than the surrounding CGM. } initial disk gas mass & $7.0\times 10^{9}\ \Msun$\\
$Z_{ISM}$ & initial disk metallicity & 2.3$\ \mathrm{Z_\odot}$\\
$R_{\rm s}$ & disk radial scale height & 3.5~kpc\\
$z_{\rm s}$ & disk vertical scale height & 0.325~kpc\\
$M_{CGM}$ & initial CGM mass (fiducial) & $2.3 \times 10^{10}\ \Msun$\\ 
$Z_{CGM}$ & initial CGM metallicity & 0.3$\ \mathrm{Z_\odot}$\\
\tableline
\end{tabular}
\caption{\label{tab:ICs} Important simulation parameters. See text for references and derivations.}
\end{table}

\subsubsection{Gaseous Disk} \label{sec:disk-ICs}

The gas of the disk follows the softened profile of \citet{tonnesen09-GasStrippingSimulated}:
\begin{equation}\label{eq:disk}
    \rho(R, z) = \frac{M_{\rm ISM}}{8\pi R_{\rm s}^2 z_{\rm s}}
        \sech\left(\frac{R}{R_{\rm s}}\right)\sech\left(\frac{z}{z_{\rm s}}\right),
\end{equation}
where $R$ is the cylindrical radius, and $R_{\rm s} = 3.5$~kpc and $z_{\rm s} = 0.325$~kpc are the scale heights of the gas. For $R > 24\ \mathrm{kpc}$, Equation \ref{eq:disk} is multiplied by a smoothing factor of 
\begin{equation} \label{eq:disk_mod}
    0.5\left[1+\cos\left(\pi\frac{R-24\ \mathrm{kpc}}{7.2\ \mathrm{kpc}}\right)\right] 
\end{equation}
that tapers the radial edges of the disk. The disk gas is given a circular velocity prescribed by the combination of our NFW and stellar potentials, with $v_\phi = \sqrt{R \cdot g(r,R,z)}$.

The disk is set to an initial uniform temperature of $10^5$~K. In reality, galaxy disks are usually at $10^4$~K or below, but gas cools very efficiently around $10^5$~K. Our disk is initialized at this higher temperature so that gas would be well above our temperature threshold for star formation ($3\times 10^3$~K; see Section \ref{sec:SF-FB}) but also rapidly cool. The variation in disk density means that the temperature does not uniformly, which helps spread out the initial burst of star formation across time. 

The initial conditions for the disk and CGM are blended together based on density; if $\rho_\mathrm{disk}(r) > \rho_\mathrm{CGM}(r)$ in a cell, that cell is considered part of the disk. Otherwise, it is initialized to be part of the CGM.

\subsubsection{Circumgalactic Medium} \label{sec:CGM-ICs}

The density and temperature structure of the CGM is set following \citet{voit19-AmbientColumnDensities}, who describes an entropy profile that precipitation-regulated galaxies in NFW halos would follow under the assumption of hydrostatic equilibrium. This assumption is reasonable for a $M_\mathrm{200} = 10^{12}\ \Msun$ halo \citep{oppenheimer18-DeviationsHydrostaticEquilibrium, lochhaas20-PropertiesSimulatedCircumgalactic}. The profile therein assumes the median $t_\mathrm{c}/t_\mathrm{ff} = 10$, but we make $\tau \equiv t_\mathrm{c}/t_\mathrm{ff}$ a free parameter in our model of the entropy profile, so that 
\begin{align} \label{eq:entropy}
    K(r)& =
        (39\ \mathrm{keV\ cm^2})v_{200}^2
        \left(\frac{r}{r_{200}}\right)^{1.1}  \nonumber \\ 
    &+ (2\mu m_p)^{1/3} 
        \left[\frac{2 n_i \tau}{3n} r \Lambda\left(2T_\phi(r), Z_{\rm CGM}\right) \right]^{2/3},
\end{align}
for entropy quantified in terms of $K \equiv kT n_e^{-2/3}$. This way we can test halos that deviate from the apparent precipitation limit of $\tau \approx 10$. We assume $n_e = n_i$ such that $2n_i/n = 1$. The initial metallicity is $Z_{\rm CGM} = 0.3$. The term $\Lambda$ is the cooling rate as a function of of temperature and metallicity, which we calculate with the Grackle chemistry and cooling library \citep{smith17-GRACKLEChemistryCooling}. The term $T_\phi(r)$ is the gravitational temperature $kT_\phi(r) = \mu m_p v_{\rm c}^2(r)/2$. 

To calculate the initial density and temperature profiles of the CGM, we combine Equation \ref{eq:entropy} with $dP/dr$ from hydrostatic equilibrium. We adopt the boundary condition $kT(r_{200}) = 0.25 \mu m_p v^2_{c,\mathrm{max}}$ suggested by \citet{voit19-AmbientColumnDensities}. Beyond $r_{200}$, the temperature begins to plummet dramatically. To mitigate this, we abandon a functional form for the entropy and instead fix $d\log P/d\log r$ to its value just inside $r_{200}$. We then set the temperature with a sigmoid function that smoothly blends $T(r_{200})$ with a floor of $4 \times 10^4$~K. The resulting mass-weighted profiles for the initial density, temperature, entropy, pressure, cooling time, and enclosed mass are shown in Figure \ref{fig:ics}. For $\tau=10$, the actual cooling time deviates slightly from the target value, falling between 7--10$t_{\rm ff}$ within $\sim0.5r_{200}$ and rises to $\sim20 t_{\rm ff}$ at the virial radius.

For simplicity, we use a modified NFW profile to calculate the circular velocity $v_{\rm c}$ used in Equation \ref{eq:entropy}. This modification approximates the presence of a galactic disk at the center of the potential well: $v_{\rm c}^2(r) = v_{\rm c,max}^2$ for $r \le 2.163 r_s$, where $r_s$ is the scale radius of the NFW profile, and
\begin{equation}
    v_{\rm c}^2(r) = 
    v_{\rm c,max}^2 \cdot 4.625\left[\frac{\ln(1+r/r_s)}{r/r_s}-\frac{1}{1+r/r_s}\right]
\end{equation}
otherwise. Using this modified NFW profile to represent the inner contents of the halo is slightly incongruous with the construction of our gas disk and static stellar potential; however, the modified NFW profile is spherically symmetric, making it easier to calculate the CGM's density and temperature profiles. This modified NFW profile is only used to set up the initial CGM conditions, while the unmodified NFW potential, the \citet{miyamoto75-ThreedimensionalModelsDistribution} stellar potential, and the gaseous disk's own gravity are what are actually applied to the simulation during execution. Any incongruities introduced into the initial hydrostatic equilibrium through our use of two slightly differing potentials will quickly fade as the simulation evolves.

We give the CGM an initial azimuthal velocity, the strength of which is determined by
\begin{equation} \label{eq:rot}
    v_\phi(r,\theta) = v_0 \sin^2\theta \left(\frac{r}{r_0}\right)^\beta.
\end{equation}
Here, $r$ is the spherical radius and $\theta$ is the polar angle. This function replaces the disk's Keplerian velocity profile wherever the CGM density dominates, as described above. We do not consider Equation \ref{eq:rot} when placing the CGM in hydrostatic equilibrium, meaning the CGM may be slightly oversupported against gravity in the cylindrically radial direction. We use \citet{hodges-kluck16-RotationHotGas} to choose $v_0=180$~km/s and $r_0=10$~kpc, which is roughly the radius where their halo model's specific angular momentum matches that of the disk (see their Figure 5). It should be noted that \citet{hodges-kluck16-RotationHotGas} adopt a constant $v_\phi$ with their model, which they assume is reasonable within $\sim 50$~kpc. We leave $\beta$ as a free parameter; for our fiducial simulation, $\beta=-1/2$ (see Table \ref{tab:sims}).

\subsection{Star Formation \& Feedback} \label{sec:SF-FB}

Our star formation and feedback algorithms are modified from the implementation of \citet{cen92-GalaxyFormationPhysical} as described in Section 2.1 of \citet{oh20-CalibrationStarFormation}. We require that gas in a grid cell have 
\begin{enumerate}
    \item $\nabla \cdot v < 0$
    \item either $t_{\rm cool} \le t_{\rm dyn}$ or $T < 3\times 10^3$~K
    \item $n \ge 10$~particles/cm$^{-3}$
    \item $m_\ast \equiv f_\ast m_{\rm cell} \frac{\Delta t}{t_{\rm dyn}} > 10^4\ \Msun$
\end{enumerate}
where the local gas dynamical time is $t_{\rm dyn} = \sqrt{ 3\pi/(32 G \rho)}$. If all these criteria are met, a ``star particle'' is created representing a population of individual stars with total mass $m_\ast$. An equivalent amount of gas is also removed from the host cell, and the particle is given a velocity such that momentum is conserved. The minimum stellar mass of $10^4\ \Msun$ is chosen as a balance between resolving the star formation with more particles and the computational expense of tracking and managing these particles. As in \citet{smith11-NatureWarmHot}, we ignore the Jeans instability criterion as it is always met by the star forming gas in our simulations. We set $f_\ast = 0.2$ and impose a minimum dynamical time of one million years.

Stellar feedback proceeds as described in Section 2.2 of \citet{oh20-CalibrationStarFormation} and is the same algorithm as used in \citet[]{peeples19-FiguringOutGas}. Though a star particle is formed immediately, for the sake of stellar feedback, the accumulation of stellar mass is assumed to be a drawn-out process that peaks in efficiency after one dynamical time \citep[][Equation 3]{oh20-CalibrationStarFormation}. The amount of mass, momentum, and energy that are returned in timestep $\Delta t$ is then tied to this extended star formation model through the mass of stars $\delta{M}_{\rm SF}$ that would form in that timestep; e.g.,
\begin{equation} \label{eq:fb}
    \Delta E = \epsilon_{\rm FB} \cdot c^2 \delta{M}_{\rm SF}
\end{equation}
We adopt the same efficiencies as \citet{oh20-CalibrationStarFormation} for the returned fraction of total and metal masses. Energy, momentum, mass, and metals are then deposited into the cube of 27 cells centered on the star particle's host cell.

It should be noted that there are no star particles present in the simulation initial conditions. This means that there are no pre-existing stellar populations affecting the gas in the earliest moments of the simulation. This artificiality has interesting consequences that are discussed in the following subsection.

\subsubsection{Ramping Stellar Feedback Efficiency} \label{sec:FB-ramp}

One of the consequences of idealized, isolated galaxy simulations like ours is the behavior of the cold gas disk at early times. The dense gas of the idealized disk is rotationally supported in the radial direction, but does not have sufficient support against gravity in the vertical direction. It therefore immediately starts to compress as the simulation begins to evolve, allowing stars to form in a large volume of the disk as the density rises. This leads to a large burst of star formation and feedback at very early times, which disrupts the CGM before it can cool significantly. This means the CGM has no time in which to influence the disk through accretion. Moreoever, feedback raises cooling times and evacuates the gas above and below the disk by successive supernova-driven shocks, preventing the CGM from interacting with the disk on a reasonable timescale. The previous isolated galaxy simulations listed in Section \ref{sec:intro} did not have this problem because they either lacked a significant CGM or did not explicitly model the star-forming disk.

Galaxies in the real universe, as well as in cosmological simulations, are slowly built up over time. They do not ``begin'' as a $\sim 7 \times 10^{9}\ \Msun$ disk of cold, dense gas. This assembly history means there have been successive generations of both stars and stellar feedback. Real galaxies will have effects such as stellar velocity dispersions and ISM turbulence that support against vertical collapse. Additionally, the CGM's dynamical and thermodynamic structure will be impacted by the accretion of matter over time and by historical outflows.

In spite of its artificiality, the value of an idealized galaxy system is that it provides a controlled environment in which to explore the interactions of the CGM and star formation. Ideally the modelled system can quickly ``relax'' into a state where the oversimplification of the initial conditions has been erased. We could minimize the collapse of the disk by adding, for instance, an initial stellar distribution (so that feedback is active at the very start of the simulation to limit the burst of star formation) or a 3D velocity dispersion; however, both of these require multiple further parameter choices to specify their distributions.

Instead, we choose to modulate the \textit{efficiency} of stellar feedback, $\epsilon_{\rm FB}$ (Equation \ref{eq:fb}). This dimensionless parameter is linearly increased between $t=1$~Gyr and $t=2$~Gyr from $\epsilon_{\rm FB} = 5 \times 10^{-8}$ to $5 \times 10^{-6}$. Our final $\epsilon_{\rm FB}$ is 0.5~dex lower than both the cosmological simulations of \citep{oh20-CalibrationStarFormation} and FOGGIE \citep{peeples19-FiguringOutGas}. The efficiency ramp begins at 1~Gyr of simulation time in order to give CGM gas with $t_{\rm cool} \lesssim 1$~Gyr time to interact with the disk before being disrupted by feedback.

\subsection{Resolution} \label{sec:amr}

Enzo uses block-structured Cartesian Adaptive Mesh Refinement \citep[AMR;][]{berger89-LocalAdaptiveMesh} to control the resolution of its grid cells. The base grid of our simulation is $128^3$ cells (12.8~kpc per cell side) and the resolution of a region is refined by a factor of two if the cell exceeds a mass threshold of $2.67\times 10^5\ {\rm M_\odot} \times 2^{(-0.5 l)}$, where $l$ is the zero-based level index. The mass threshold for refinement therefore decreases with level in a super-Lagrangian way, preventing excessive refinement on the lowest levels which would increase computational cost. The highest levels are concentrated in the galactic disk, which, while not the focus of this study, is important to resolve for the purposes of star formation.

\begin{figure}
    \centering
    \includegraphics[width=\linewidth]{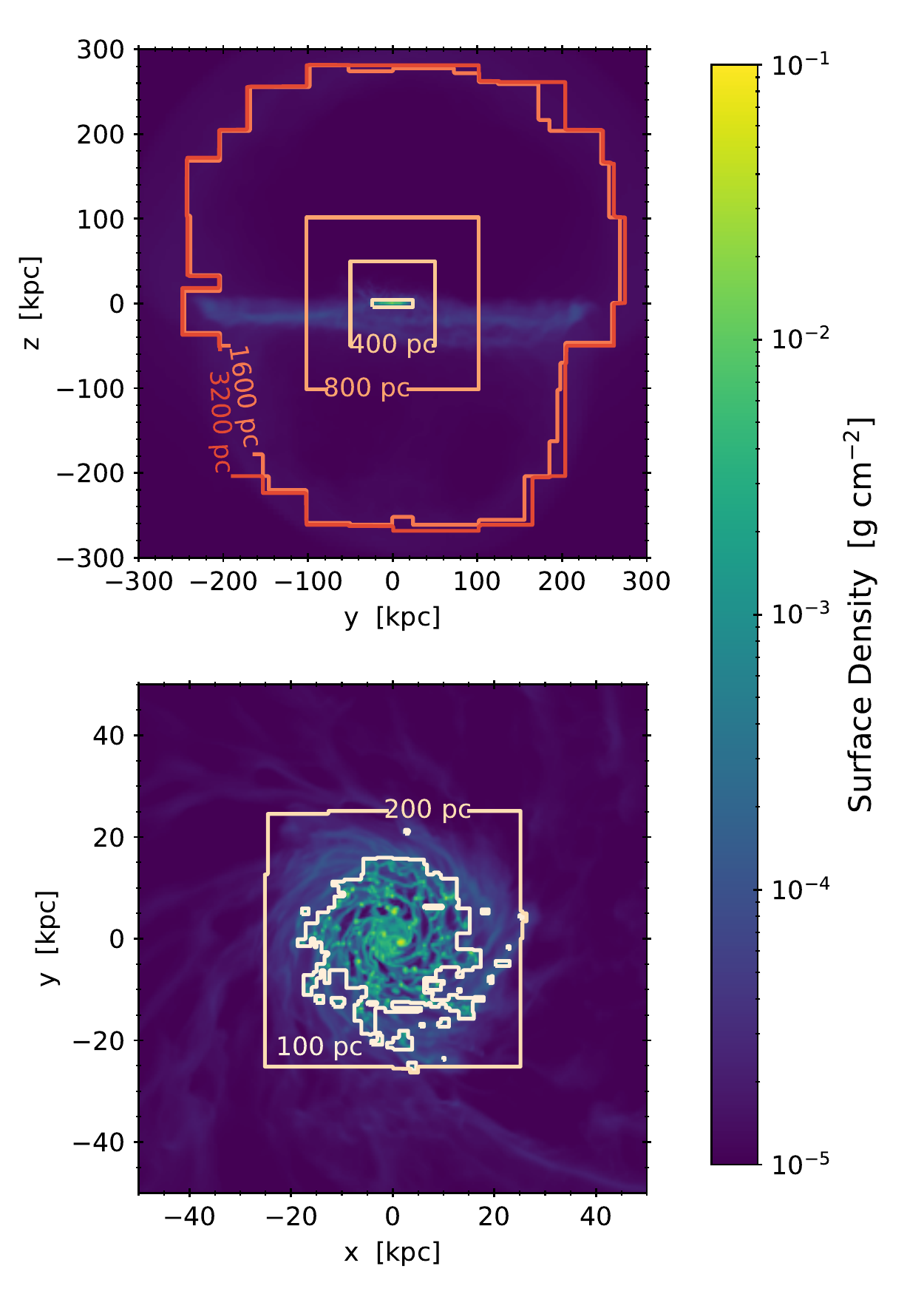}
    \caption{Demonstration of our AMR scheme with edge-on (top) and zoomed face-on (bottom) density projections at $t=3$~Gyr. Orange contours enclose regions of uniform resolution at the center of the projection. Our refinement scheme defines nested boxes that must be at a fixed minimum resolution, but cells are allowed to refine beyond this minimum if they exceed a mass threshold. This behavior can be seen with the {3,200}, {1,600}, and 200~pc contours. The 100~pc contour is based exclusively on mass refinement and is the highest resolution allowed.}
    \label{fig:amr_onecol}
\end{figure}

Given the low density of the CGM, a mass-based refinement criterion is not enough for the CGM gas to become well-resolved. The importance of good CGM resolution has been demonstrated by \citet{hummels19-ImpactEnhancedHalo}, \citet{peeples19-FiguringOutGas}, \citet[]{suresh19-ZoomingAccretionII}, and \citet{vandevoort19-CosmologicalSimulationsCircumgalactic}. Smaller clouds are allowed to develop with higher CGM resolution, and turbulent structures are resolved to smaller scales. We therefore also define six nested rectangular regions of fixed minimum resolution (the first four of which are cubic). Cells within these regions are allowed to refine further based on mass. An example of this is shown in Figure \ref{fig:amr_onecol}. We allow up to seven levels of refinement (a minimum spatial resolution of 100~pc).

\subsection{Simulation Variants} \label{sec:variants}

\begin{table}[t]
\centering
\begin{tabular}{lrrr}
\toprule
Name & FB & $t_\mathrm{c}/t_\mathrm{ff}$ & $\beta$ \\
\tableline
\textsc{Fiducial} & Y & 10 & -1/2 \\
\textsc{CoolFlow} & \textbf{N}  & 10 & -1/2\\
\textsc{LowRatio} & Y & \textbf{5} & -1/2 \\
\textsc{HighRatio} & Y & \textbf{20} & -1/2\\
\textsc{LinRot} & Y & 10 & \textbf{-1} \\
\textsc{NoRot} & Y & 10 & \textbf{N/A} \\
\tableline
\end{tabular}
\caption{\label{tab:sims} Our set of simulations and their varied parameters. Variations from the fiducial model are highlighted in bold. The parameter $\beta$ refers to the power law index of the initial azimuthal rotation profile.
}
\end{table}

The goal of our simulations is not only to test if we can create a (precipitating) self-regulating system, but also to explore the robustness of self-regulation. To that end, we explore five variations on our fiducial simulation, as laid out in Table \ref{tab:sims}. The first variation, \textsc{CoolFlow}, is identical to the fiducial run in its initial conditions but has stellar feedback completely disabled (star formation is still allowed in order to remove cold gas and preserve numerical stability). In this way, it functions as a control to demonstrate the importance of feedback. 

The next two variants, \textsc{LowRatio} and \textsc{HighRatio}, modify the precipitation limit parameter $\tau$ in Equation~\ref{eq:entropy}. Their values of $\tau$ are chosen to be the approximate lower and upper bounds experienced, on average, by a precipitating system \citep{voit17-GlobalModelCircumgalactic, voit18-RoleTurbulenceCircumgalactic,voit21-GraphicalInterpretationCircumgalactic}. We note that, because the CGM's radial density profile is defined by the initial $\tau$ through Equation \ref{eq:entropy}, these variants start with CGM masses different by about a factor of 2. Compared to the fiducial CGM mass given in Table \ref{tab:ICs}, the \textsc{HighRatio} variant starts with $1.3\times10^{10}\ \Msun$ of CGM gas and the \textsc{LowRatio} with $3.9\times10^{10}\ \Msun$.

Though the CGM of galaxies likely rotates \citep{hodges-kluck16-RotationHotGas, defelippis20-AngularMomentumCircumgalactic}, the shape of its rotation profile is not well-constrained. The \textsc{LinRot} variation modifies the initial rotation profile of the CGM, changing the index $\beta$ in Equation \ref{eq:rot}, while \textsc{NoRot} is a control for the impact of CGM rotation on the evolution of the system. This latter variant has no initial rotation in the CGM, but still maintains Keplerian rotation in the disk.

\section{The Galactic Disk}\label{sec:disk}

We start by looking at the gas content of the galactic disk and its star formation. In later sections, we will look in detail at the exchange of gas between the disk and CGM, but the effects of CGM accretion are readily apparent within the disk.

\subsection{Appearance}\label{sec:disk-pics}

\begin{figure}
    \centering
    \includegraphics[width=\columnwidth]{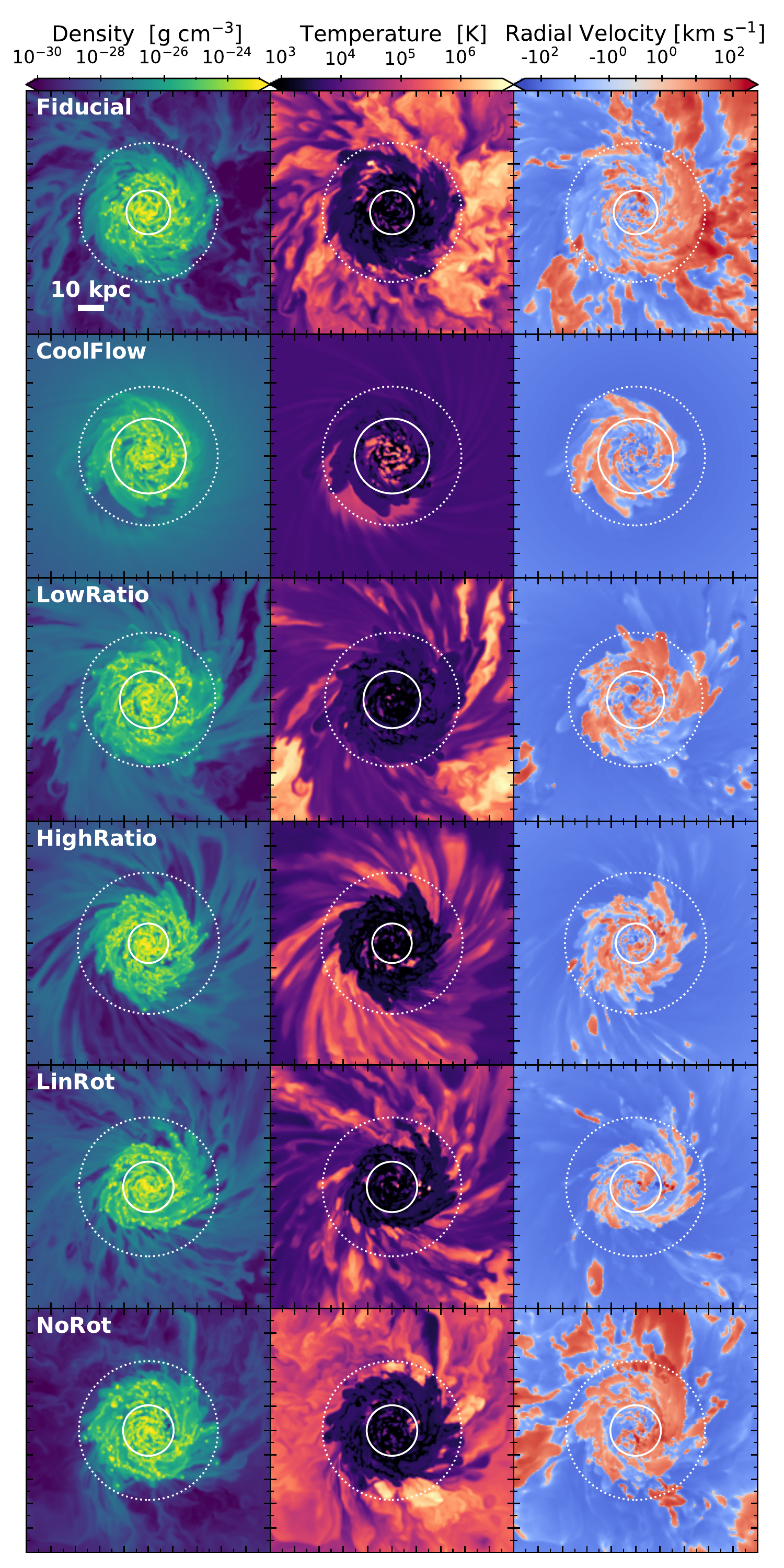}
    \caption{Slices of density, temperature, and radial velocity through the disk midplane in each simulation variant at 3~Gyr. Cells are 100~pc thick in the center of the galaxy, 200~pc thick out to $\pm 25$~kpc, and 400~pc thick out to the edge of the field of view at $\pm 50$~kpc from the center.
    Solid white circles show the average radius within which stars form over 2--4~Gyr. 
    Dotted white circles mark the extent of the initial gas disk ($\sim 28$~kpc). Note that the colormap for radial velocity covers $\pm3\times10^3$~km/s.
    }
    \label{fig:face_comp}
\end{figure}

\begin{table}
\centering
\begin{tabular}{lrr}
\toprule
Variant & Mean (kpc) & Std Dev (kpc) \\ 
\tableline
\textsc{Cflow}   &   15.4 & 2.1 \\ 
\textsc{LowRatio}  & 11.7 & 4.2 \\ 
\textsc{LinRot}  &   10.4 & 3.7 \\ 
\textsc{NoRot}   &   10.4 & 5.1 \\ 
\textsc{Fid}     &    9.0 & 3.9 \\ 
\textsc{HighRatio} &  8.1 & 3.7 \\ 
\tableline
\end{tabular}
\caption{\label{tab:SFradii} Mean and standard deviation of the maximum star formation radius over 2--4~Gyr of simulation time, in kpc. The maximum SF radius is defined as the largest cylindrical radius at which a star formed in the last 50~Myr (which is the time between simulation outputs).
}
\end{table}

Figure \ref{fig:face_comp} compares the disks of our simulation variants at a simulation time of 3~Gyr. We show the density, temperature, and radial velocity within a slice through the disk midplane. The solid white circles represent the average radius within which stars form, which is determined as follows: for each simulation snapshot ($\Delta t=50$~Myr), we determine the maximum cylindrical radius of star particles formed since the last snapshot. The average of these maxima over $t=2$--4~Gyr give the white circles in Figure \ref{fig:face_comp}. The numerical values of these radii, as well as their standard deviations over time, are reported in Table \ref{tab:SFradii}. The dotted circles show the edge of each variant's initial gas disk. Due to differences in initial CGM structure and the way the disk and CGM are blended together (see Section \ref{sec:CGM-ICs}), the \textsc{LowRatio} and \textsc{HighRatio} simulations have slightly different initial radii than the other variants, but all are around 28~kpc.

From these slices we see that the gas in the disk is around $10^4$~K on average, reaching $10^3$~K and lower in the densest spiral arms and near the center. Star formation is most prevalent in the very center of the disk. There are hotspots of $\sim10^5$~K gas within the star-forming center of each variant's disk that are the result of feedback. The exception is the \textsc{CoolFlow} simulation, which has more prominent hotspots but no feedback. For this variant, the gas heating is due to a ``pile-up'' of accreting gas. This pile-up will be discussed further in Section \ref{sec:edge-pics}. 

The radial velocity slices show us that the hottest gas at $T\sim10^6$~K is not strictly outflowing. Clear examples of this can be seen in the right side of the fiducial slices, in the bottom corner of the \textsc{LowRatio} images, and throughout the \textsc{NoRot} slices. The fiducial simulation, in particular, shows evidence of a radial velocity gradiant across a hot, low density cloud. This, along with the highly variable gas temperature outside the galactic disk, suggests that hot gas is mixing with cooler gas in the disk midplane. The prevalence of radial inflow in hot gas also suggests that gas is cooling as it moves towards the disk. Though highlighted in Figure \ref{fig:face_comp} by \textsc{LowRatio} and \textsc{NoRot}, cooling inflow is present for $t>2$~Gyr in all variations.

The size of the disk's star forming region appears impacted by the CGM's $\tctff$ properties, as seen from the solid white circles in Figure \ref{fig:face_comp} and the values in Table \ref{tab:SFradii}. For both of these, we use the average maximum radius within which stars form as a measure of the star forming region. The \textsc{LowRatio} variant has an SF region that is statistically larger than the fiducial and the \textsc{HighRatio} variants (adopting $\alpha=0.05$ or a 95\% confidence interval), though we note that the fiducial's region is not signficantly larger than \textsc{HighRatio}'s. This suggests a small dependence on the simulation's initial $\tctff$ ratio. Moreover, the fiducial, \textsc{LinRot}, and \textsc{NoRot} simulations are statistically indistinguishable when tested against each other\footnote{
A slight difference between the fiducial, \textsc{LinRot}, and \textsc{NoRot} simulations appears when testing each of these simulations against other variants. While the \textsc{LowRatio} simulation has a statistically larger star forming region than the fiducial, it fails to test as significantly larger than the \textsc{LinRot} and \textsc{NoRot} simulations. Additionally, while the fiducial simulation does not test as significantly larger than the \textsc{HighRatio} simulation, the \textsc{LinRot} and \textsc{NoRot} variants do.
}. These simulations all have the same initial $\tctff$ ratio. The \textsc{CoolFlow} simulation has a star forming region that is statistically significantly larger than all the other simulation variants, but this follows from its lack of feedback and aggressive star formation rather than a difference in its CGM.

\subsection{Mass Growth}\label{sec:disk-mass}

\begin{figure*}
    \centering
    \includegraphics[width=\textwidth]{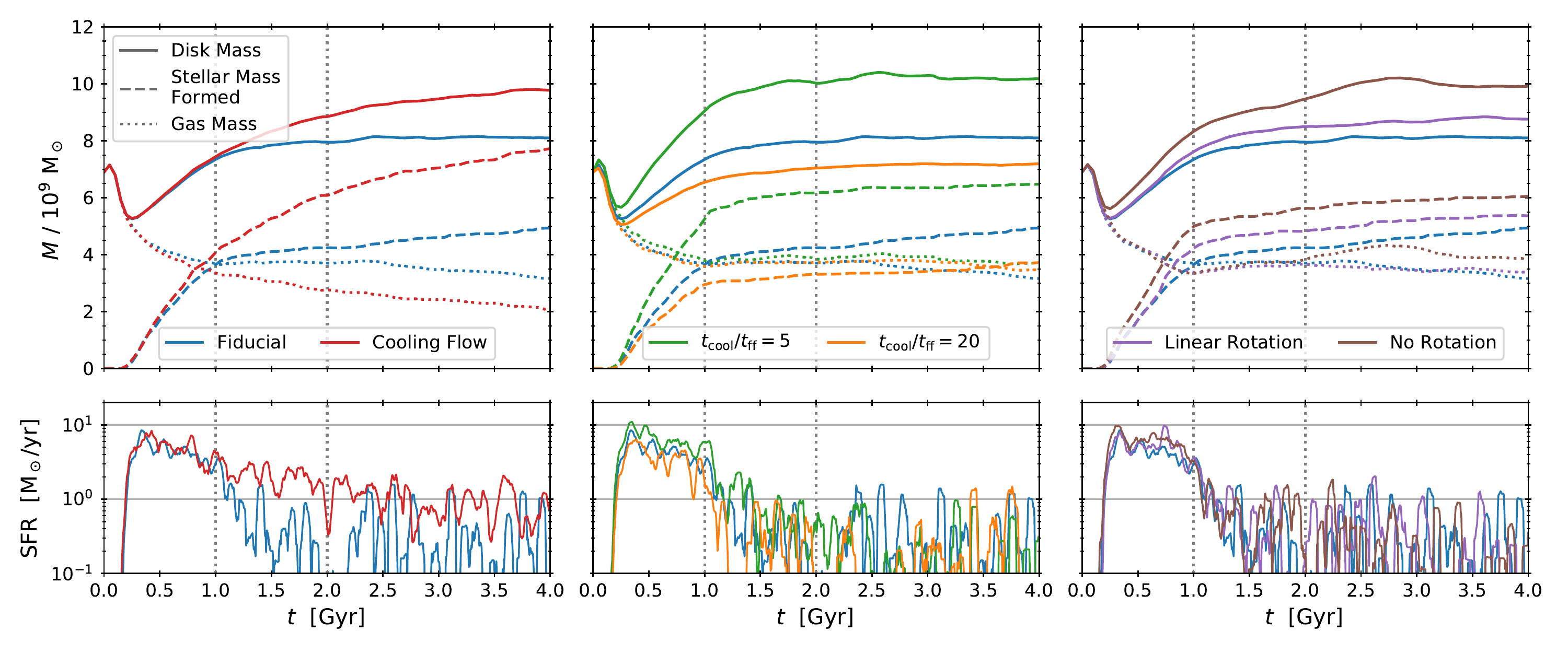}
    \caption{Disk mass (top) and SFR (bottom) as a function of simulation time. Disk mass (solid) is broken into gas (dotted) and stellar (dashed) components. The stellar mass reflects the cumulative mass of stars formed and neglects stellar mass loss. To improve clarity, the SFR is shown as the rolling mean with a 60~Myr rectangular window. The region between the two dotted grey lines is when the stellar feedback efficiency is ramped (Section \ref{sec:FB-ramp}). From left to right, the \textsc{Fiducial} model (blue) is compared with \textsc{CoolFlow}, \textsc{LowRatio} and \textsc{HighRatio}, and \textsc{LinRot} and \textsc{NoRot}. Compared to the \textsc{CoolFlow} variant, the mass growth in each component remains fairly stable after the onset of feedback at 1~Gyr.
    The final stellar mass varies for each variant, along with the early-time ($t<1$~Gyr) SFR and gas consumption.}
    \label{fig:mass_ev}
\end{figure*}

\begin{figure}
    \centering
    \includegraphics[width=\linewidth]{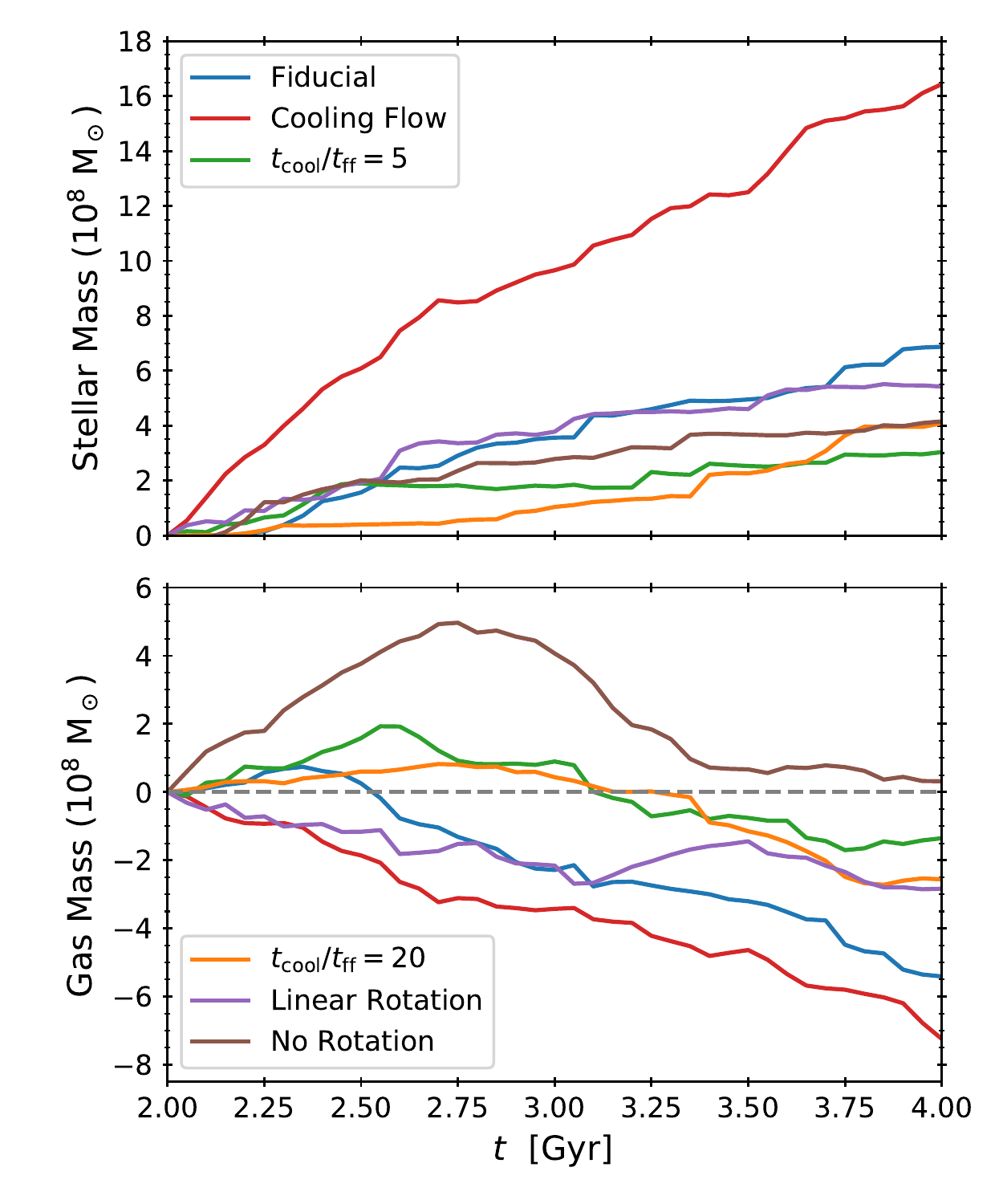}
    \caption{Mass of the stellar (top) and gaseous (bottom) disk components relative to $t=2$~Gyr. The legend is split between both panels. The stellar mass grows similarly among the non-\textsc{CoolFlow} runs, but the gas mass varies widely.
    }
    \label{fig:mass_ev_late}
\end{figure}

Figure \ref{fig:mass_ev} shows the disk mass and star formation rate over the full simulation time (4~Gyr) for each variant. The disk mass is split into gaseous and stellar components. The disk gas is defined by a cylindrical region with $R\approx5.7R_{\rm s}=20$~kpc and thickness $8z_{\rm s}=2.6$~kpc.
This cutoff in cylindrical radius is somewhat arbitrary, but corresponds roughly to a transition point in the profiles of density and temperature that is visible from 2--4~Gyr. The disk's stellar mass is defined as the total mass of all stars formed, and neglects stellar mass loss. Including stellar mass loss lowers the stellar masses at $t>1$~Gyr by $\sim 1.5\times10^9\ \Msun$. The accompanying SFR has been smoothed using a trailing moving average over a $\sigma=60$~Myr rectangular window. Vertical dotted lines denote the beginning and end of the feedback efficiency ramp laid out in Section \ref{sec:FB-ramp}. The fiducial simulation is recreated in all three panels.

In the leftmost panel, the fiducial simulation is compared to the \textsc{CoolFlow} run. For the fiducial model, the growth in stellar mass slows past 1~Gyr, when the stellar feedback efficiency begins to ramp. In the \textsc{CoolFlow} model, however, the stellar mass continues to grow significantly as the simulation progresses, depleting the disk's gas mass more appreciably than in the fiducial variant. The SFR is also generally higher in the \textsc{CoolFlow} simulation. This behavior is expected as the \textsc{CoolFlow} model has stellar feedback disabled. This also indicates the flattening of the fiducial model's stellar growth is due to feedback, which is confirmation of an expected result.

In the middle panel, variations in the initial $\tctff$ ratio are explored by comparing the fiducial simulation to the \textsc{LowRatio} and \textsc{HighRatio} runs. The rightmost panel explores the rotation variants \textsc{LinRot} and \textsc{NoRot}. All of the variants with stellar feedback demonstrate remarkably similar gas mass curves. Stellar growth and gas mass loss flatten out around 1~Gyr, when the feedback efficiency begins to increase. The variants are primarily distinguished by the total amount of stellar mass they form, with the largest difference observed between the $\tctff$ variants. 

For the first $\sim0.7$~Gyr, the fiducial and \textsc{LinRot} simulations have the same stellar mass curves. This is also true for the fiducial and \textsc{CoolFlow} simulations before $\sim0.5$~Gyr. Since these simulations all start with an identical initial disk, deviations in stellar mass can be attributed to gas accretion from the CGM. The \textsc{NoRot} variant has little apparent overlap in stellar mass growth with the fiducial simulation, indicating an earlier deviation in the amount of CGM accretion. We attribute \textsc{NoRot}'s early differentiation to the lack of CGM angular momentum, which should make it easier for gas to settle onto the disk and form stars. The \textsc{LowRatio} and \textsc{HighRatio} simulations start with slightly smaller/larger initial disk radii, respectively, due to how the disk and CGM are blended (Section \ref{sec:disk-ICs}). This has a negligible effect on their disk mass, while their CGM masses deviate from the fiducial by about a factor of two with \textsc{LowRatio} being the most massive. Given that the \textsc{LowRatio} simulation goes on to form the most stars (of the variants with feedback), this is further evidence that CGM accretion is a major contributor to stellar growth.

The fiducial model and all other variants with $\tau=10$ start with about $8\times10^8\ \Msun$ of CGM gas with $t_{\rm cool} \le 1$~Gyr, which is the timescale on which stellar feedback is kept inefficient. The \textsc{LowRatio} variant has $\sim4\times10^9\ \Msun$ of this short-cooling time gas ($\sim5$ times greater than the fiducial), and forms the most stars before 1~Gyr. On the other hand, the \textsc{HighRatio} variant forms the fewest stars by 1~Gyr, and its CGM only has $\sim1\times10^8\ \Msun$ of gas with $t_{\rm cool} \le 1$~Gyr (a factor of 8 lower than the fiducial). These differences support the interpretation that star formation within the first gigayear is fueled not only by disk mass, but also by the accretion of gas from the CGM. Additionally, there is evidence for continued CGM inflow as the gas mass continues to increase past 1~Gyr.  The inflow of gas will be discussed more in Section \ref{sec:gas-budget}.

Though each variant experiences a different amount of star formation, particularly at early times, their gas masses remain both relatively constant and similar across variants. Additionally, the smoothed star formation rates are also quite consistent once stellar feedback reaches its full strength at $t=2$~Gyr. We attribute the constancy of the disk mass among simulation variants to the Toomre criterion \citep{toomre64-GravitationalStabilityDisk}. A mass of $\sim4 \times 10^9\ \Msun$ appears to be the value at which the disk is marginally stable against gravitational instability. As gas is added to an isothermal disk, the disk gas is able to fragment and form stars until the surface density falls back below the threshold for instability. This effect can also explain why the \textsc{LowRatio} variant experiences more star formation from 1--2~Gyr while maintaining a fairy stable gas mass in the disk. The role of the Toomre instability will be discussed more  in Section \ref{sec:toomre}.

In Figure \ref{fig:mass_ev_late} we focus on the late time ($t \geq 2$~Gyr) stellar and gas masses. These quantities are shown relative to each simulation's gas and stellar masses at $t=2$~Gyr. Once again, stellar mass loss is ignored in order to show the integrated star formation. Including stellar mass loss drops the final stellar mass increase by $\sim1\times10^8\ \Msun$ for the simulations with feedback and $\sim 4\times 10^8\ \Msun$ for the \textsc{CoolFlow} variant. 

The most dramatic stellar mass increase can again be seen in the \textsc{CoolFlow} simulation, while the variants with feedback experience more gradual increases in their stellar masses. The fiducial and \textsc{LinRot} simulations grow at a similar rate, while the \textsc{LowRatio} and \textsc{NoRot} variants ``taper off'' in their stellar mass increase after 2.5~Gyr. The \textsc{HighRatio} variant appears to be a ``late bloomer,'' with stellar mass increases smaller than the other variants with feedback until around 3.6~Gyr. All of the variants with feedback form between 3--$7\times10^8\ \Msun$ more stars after $t=2$~Gyr.

These final increases in stellar mass are partially compensated by overall decreases in gas mass, which, ignoring the \textsc{NoRot} and \textsc{CoolFlow} variants, are between 1.5 and $5.5\times10^8\ \Msun$ at $t=4$~Gyr. Each simulation variant, however, took a very different path to reach this point. Except for the \textsc{LinRot} and \textsc{CoolFlow} simulations, most variants see an increase in their disk's gas mass at some point before $\sim2.3$~Gyr. The \textsc{NoRot} simulation sees the largest increase in disk gas mass, and is the only simulation to end with more gas at $t=4$~Gyr than it had at $t=2$~Gyr. Despite this, it is not the variant with the largest increase in stellar mass. The differences in disk gas mass between the variants highlight that the story of accretion is complex. We will explore it more in Section \ref{sec:gas-budget}.

\section{Outflows and Inflows}\label{sec:inflow-outflow}

Feedback processes in the disk are able to drive large-scale galactic outflows. In our simulations, these ``feedback processes'' are limited to Type II supernovae. Indeed, self-regulation requires that inflowing gas must balance gas loss through outflows and star formation. Given that our idealized models ignore galaxy mergers, gas filaments, and other cosmological processes, their evolution is a story of inflowing and outflowing CGM gas. Therefore, in this section we examine the exchange of gas between the CGM and the disk as our simulations evolve.

\subsection{System Dynamics}\label{sec:edge-pics}

\begin{figure*}
    \centering
    \includegraphics[width=\textwidth]{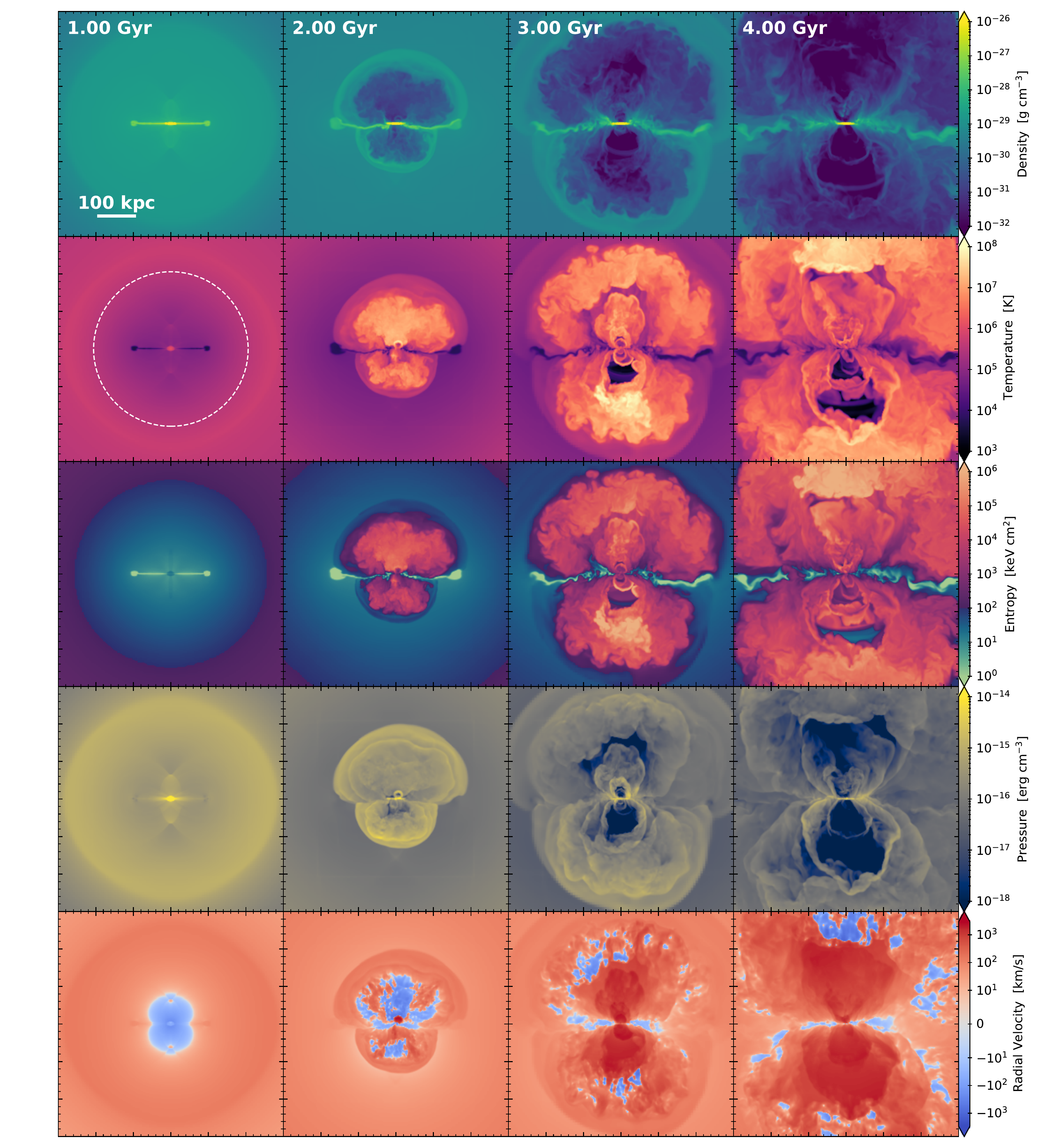}
    \caption{Average line-of-sight density, temperature, entropy, pressure, and radial velocity of the fiducial run at four different times. Averages are taken from a thin slab two radial scale heights $2R_{\rm s}=7$~kpc thick and 600~kpc square centered on the disk, which is shown edge-on. The white circle shows $r_{200} \approx 206$~kpc. The entropy color bar separates gas with $K>10^2$~keV~cm$^2$ in pink, which should not cool within a few Gyr. The efficiency of stellar feedback is ramped between 1 and 2~Gyr (Section \ref{sec:FB-ramp}).}
    \label{fig:fid_ev}
\end{figure*}

\startlongtable
\movetabledown=1.75in
\begin{figure*}[p]
\begin{rotatetable*}
    \movetabledown=1in
    \includegraphics[height={\dimexpr\textwidth-1.5\baselineskip\relax}]{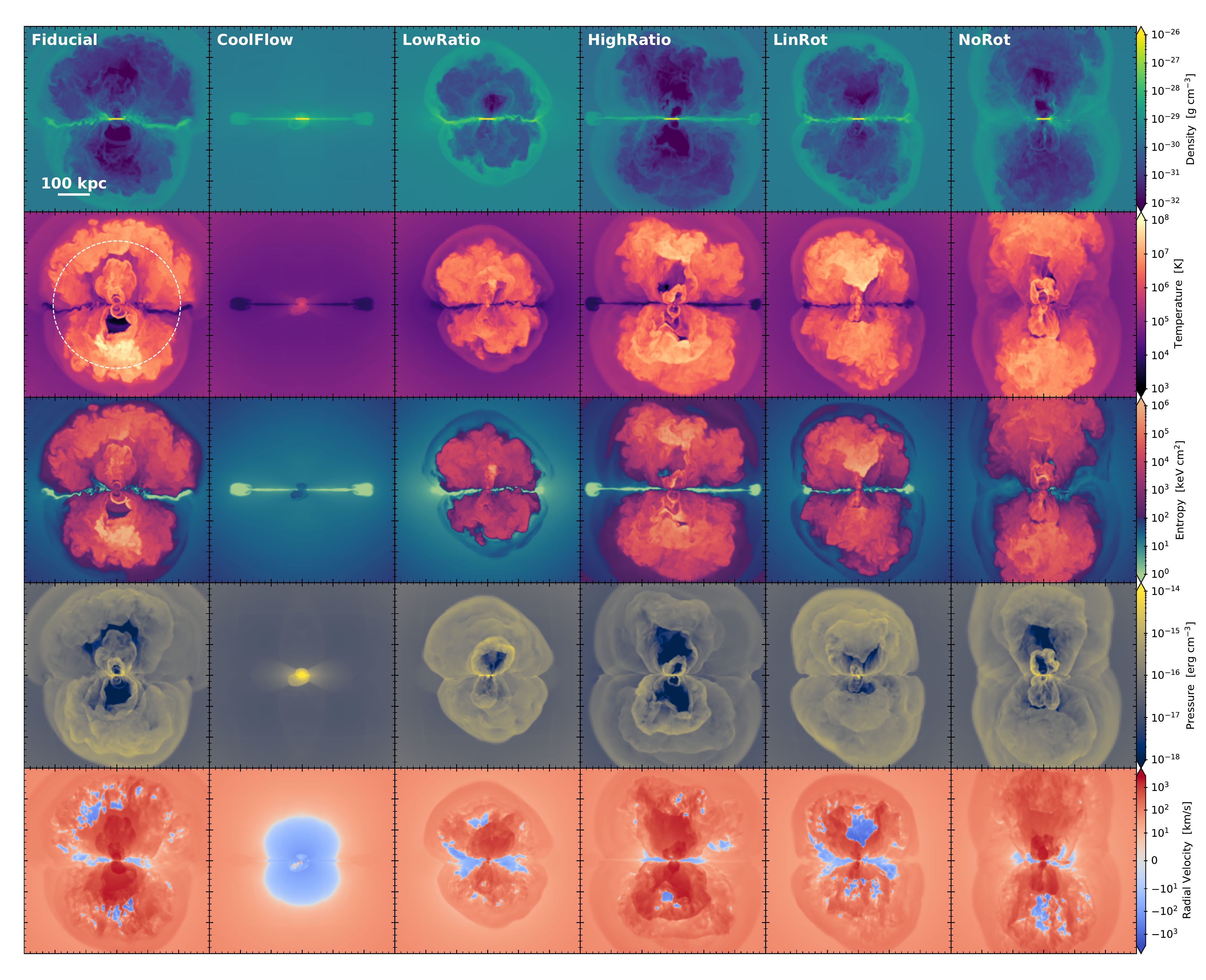}
    \caption{Same as Figure \ref{fig:fid_ev} but for each simulation variant at 3~Gyr. The leftmost column is reproduced from Figure \ref{fig:fid_ev}.}
    \label{fig:edge_comp}
\end{rotatetable*}
\end{figure*}

Figure \ref{fig:fid_ev} shows the evolution of the fiducial simulation in four thermodynamic quantities---density, temperature, entropy, and pressure---and radial velocity. Average values along the line of sight are shown at four times (in gigayear intervals) using an edge-on thin slab. These slabs are two radial scale heights ($2 R_{\rm s} = 7$~kpc) thick and 600~kpc square on their sides, centered on the disk. By averaging through a thin subdomain, we can show a more representative visualization of the domain (as compared to a slice) while not oversmoothing turbulent gas structures. The white circle denotes $r_{200}$. The ramp in stellar feedback efficiency, as described in Section \ref{sec:FB-ramp}, occurs from 1--2~Gyr.

At 1~Gyr, the initial burst of star formation begins to subside as the feedback efficiency starts to increase. The initial disk has collapsed as described in Section \ref{sec:FB-ramp}, pushing out gas in the radial direction. Sound waves were driven by the disk's initial collapse, including a spherical wave that has propagated beyond the virial radius at 1~Gyr. Its propagation has resulted in a positive radial velocity throughout the domain. Gas behind this wave has cooled from its initial temperature profile and started falling inward.
Net inward flow can be seen in blue in the bottom row of Figure \ref{fig:fid_ev} and is approximately 50 kpc in radius at $t=1$~Gyr.
Further evidence of sound waves can be seen faintly in the pressure above and below the disk. These intersecting wave patterns will eventually be wiped away by stellar feedback, which we can see starting to happen at 2~Gyr. 

At $t=2$~Gyr we see outflows have disrupted the inflowing gas above and below the disk. Some gas inside these outflows has started falling back towards the galaxy. By 3~Gyr, these outflows have expanded past the virial radius. Gas near the edge of older outflows still exhibits some inward radial motion, but is disrupted by younger outflows. The interior of these recent outflows has very low densities and pressures; lower than at 2~Gyr. At both $t=2$ and 3~Gyr, cold dense gas is feeding the disk along its midplane.

By 4~Gyr, we can see that the history of successive feedback-driven outflows has created a complex shock structure within 100--200~kpc directly above and below the disk. Visible at the very top of the projection is an inflowing front with high temperature and entropy, but very low density ($<10^{-32}$~g/cm$^3$). Indeed, there is very little gas above and below the disk. Meanwhile, accreting material continues to flow inward along the midplane. There is gas infall at large radii that is outside the disk midplane, but this is prevented from reaching the disk by younger outflows.

In Figure \ref{fig:edge_comp}, we compare the fiducial simulation at 3~Gyr to each of the variants, still using the line-of-sight average through a thin slab. With no feedback to disrupt its CGM, the \textsc{CoolFlow} variant has a rough sphere of inflowing gas around its disk. The high pressure region at the disk's center steadily grows in radius from the continually infalling gas. This increase in pressure also drives an increase in temperature. 

All other variants have outflow ``plumes'' that vary in size and shape. The \textsc{NoRot} variant has a taller, narrower outflow envelope than the fiducial and \textsc{LinRot} models, indicating that rotation in the CGM may be important for distributing outflowing material such as metals evenly through the CGM. Visible in the \textsc{LinRot} model is another high-entropy, low-density front of gas falling back towards the disk. These features develop in many of the simulations, but are prevented from reaching the disk by futher feedback.

Clearly visible in the \textsc{CoolFlow} variant, but present in all simulations except \textsc{NoRot}, is the remnant of the initial disk's radial spread due to vertical collapse (Section \ref{sec:FB-ramp}). This gas is very low entropy, and though not shown, has metallicity associated with the initial disk. 
This initial disk spreads out the least for the \textsc{NoRot} simulation (reaching about half as far as the fiducial simulation at $t=2$~Gyr), indicating the initial expansion of the disk is encouraged by the CGM's own angular momentum. For \textsc{NoRot}, this smaller spreading feature is easily disrupted by feedback, which is why this feature is not as prominent in Figure \ref{fig:edge_comp} and why hot gas appears much closer to the spiral disk in Figure \ref{fig:face_comp}.

\subsection{Gas Availability \& Accretion}\label{sec:gas-budget}

\begin{figure*}
    \centering
    \includegraphics[width=\linewidth]{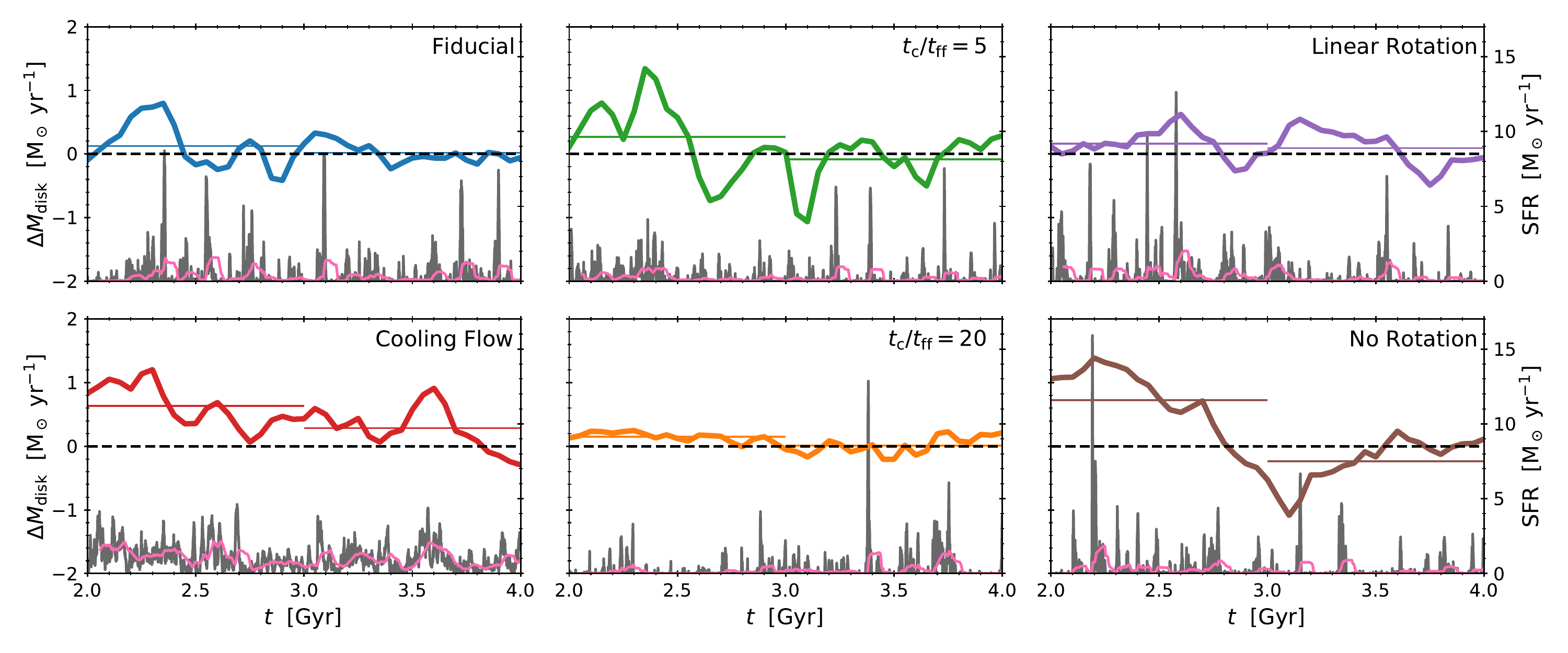}
    \caption{Net rate of change in disk mass $\dot{M}_{\rm disk}$ (left axis; thick lines) compared to the star formation rate (right axis; gray and pink). The rate of disk mass change is calculated as $\dot{M}_{\rm disk} = \dot{M}_\ast + \dot{M}_{\rm gas}$. The dashed horizontal line marks the zero-point for this quantity.
    Note that all displayed quantities are sampled at different cadences: $\Delta t = 50$~Myr for the $\dot{M}_{\rm disk}$ and $\Delta t=0.6$~Myr for the gray SFR. The 60~Myr moving-averaged SFR from Figure \ref{fig:mass_ev} is shown in pink for comparison, and the horizontal lines indicate average net $\dot{M}_{\rm disk}$ over 1~Gyr.
    }
    \label{fig:net_gas_flow}
\end{figure*}

\begin{figure*}
    \centering
    \includegraphics[width=\linewidth]{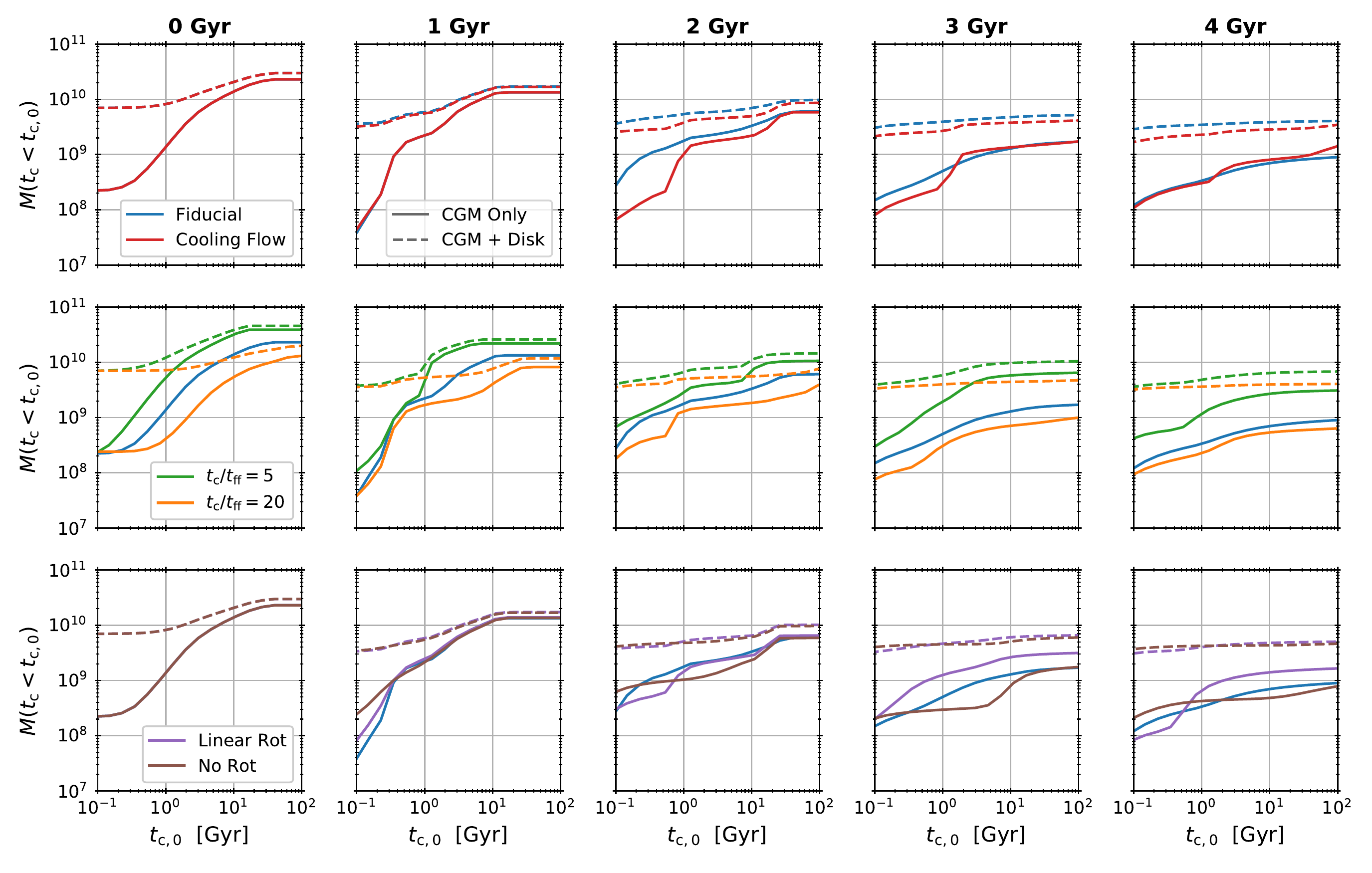}
    \caption{Cumulative mass distribution of CGM gas as a function of cooling time at five simulation times: 0, 1, 2, 3, and 4~Gyr. The CGM is defined by $r<r_{200}=206$~kpc, with a cylinder 40~kpc in diameter and $8z_{\rm s} = 2.6$~kpc thick excised to remove the disk.
    The top row compares the \textsc{Fiducial} model (blue) with the \textsc{CoolFlow} variant, the middle row with \textsc{LowRatio} (green) and \textsc{HighRatio} (orange), and the bottom row with \textsc{LinRot} (purple) and \textsc{NoRot} (brown). The initial distribution (leftmost column) is identical for all but the $\tctff$ variants.
    }
    \label{fig:tcool_dist}
\end{figure*}

\begin{figure*}
    \centering
    \includegraphics[width=\columnwidth]{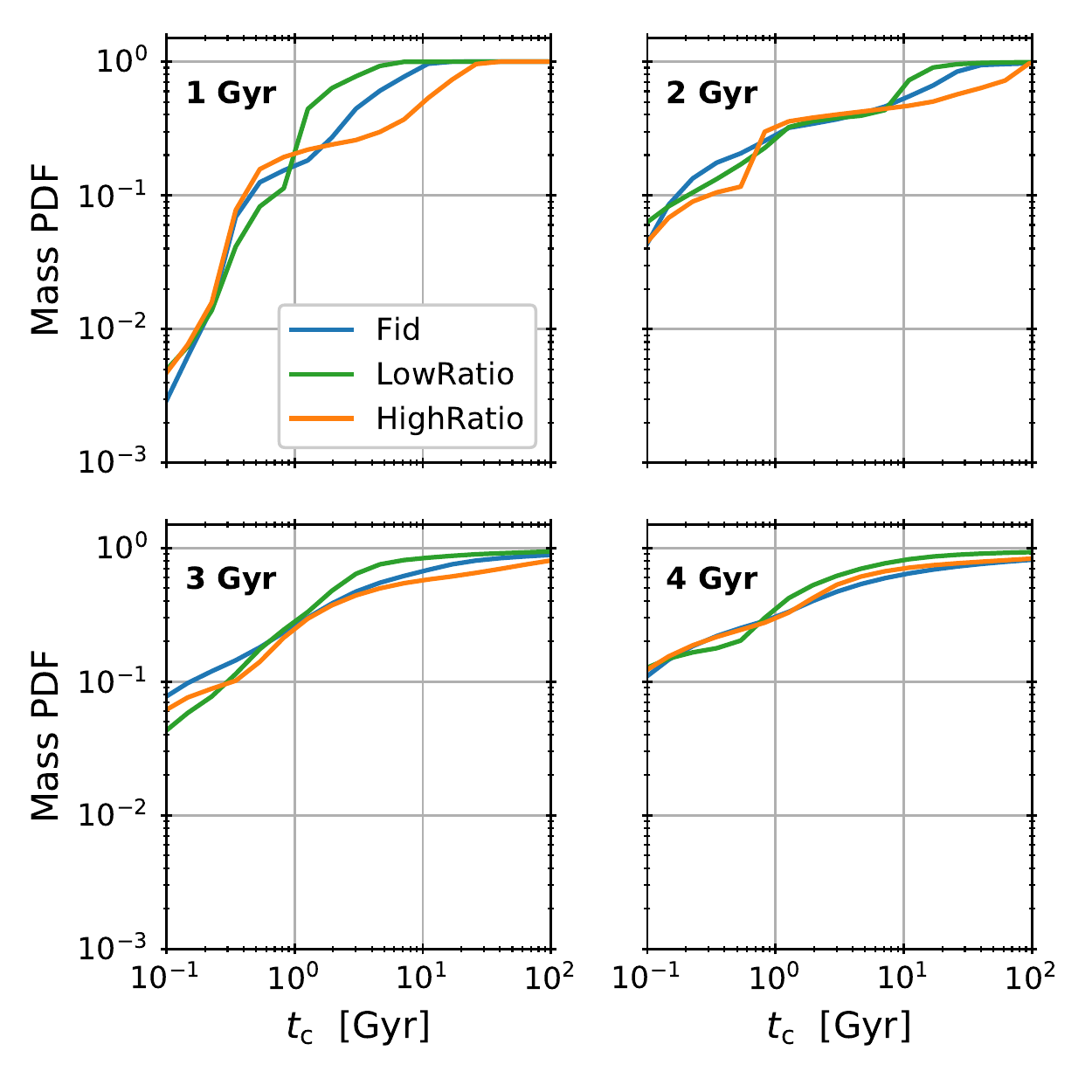}
    \includegraphics[width=\columnwidth]{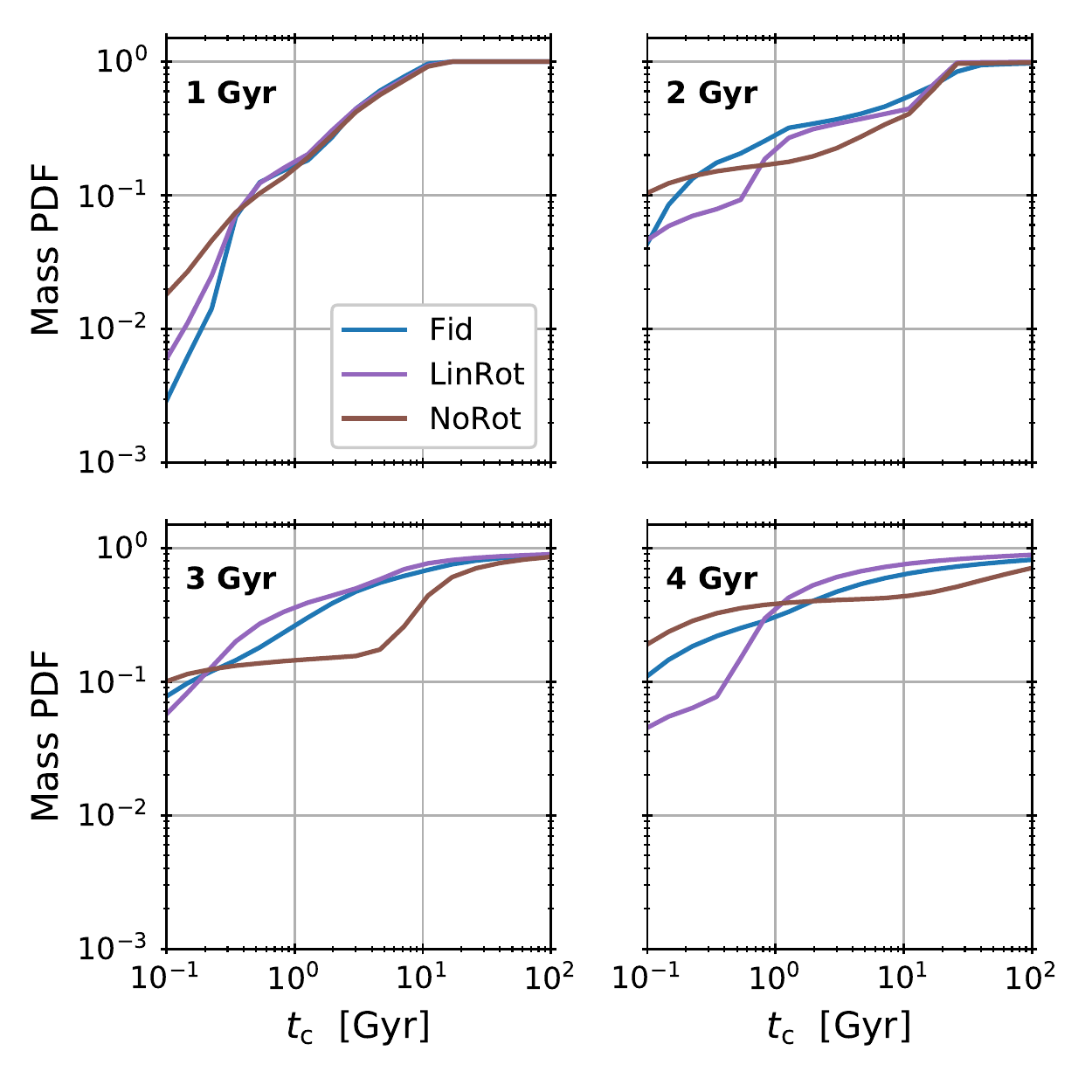}
    \caption{The same as the middle (left) and bottom (right) rows of Figure \ref{fig:tcool_dist}, but the mass distributions have been normalized in order to highlight differences in shape at each timestamp. The simulations with varying initial $\tctff$ maintain cooling time distributions with similar shapes. On the other hand, the variants with different CGM rotation profiles vary considerably from each other across time.}
    \label{fig:tcool_norm}
\end{figure*}

In our simulations, there are two major sources of gas for star formation: cold gas that was present in the disk from the initial conditions, and CGM gas that accretes onto the galaxy.  Dying stars also return gas to the disk, but their contribution over a $\lesssim 1$~Gyr timescale isnegligible in comparison to these other two sources \citep[unlike in ellipticals; see][]{voit11-FateStellarMass}. The story of gas accretion can also be thought of as the story of gas availability: what gas can cool and fall inward toward the disk? How much is gas in a rotating CGM able to shed angular momentum? In Figure \ref{fig:mass_ev}, the differences in stellar mass formed by 1~Gyr are a clear indicator that our simulation variants experience different rates of gas accretion and star formation. In Section \ref{sec:disk-mass} we noted that these differences depend primarily on the mass of CGM gas with $t_{\rm cool}<1$~Gyr and on the presence of rotation. Figure \ref{fig:mass_ev_late} suggests that the simulation variants have unique and variable patterns of gas inflow over time. In this section, we delve more deeply into the ability of the CGM to contribute gas to the disk.

Figure \ref{fig:net_gas_flow} shows the net rate of change in disk mass for $t>2$~Gyr, which is when stellar feedback is at its full efficiency. The disk is restricted to a cylinder with $R=20$~kpc and thickness $8z_{\rm s}=2.6$~kpc at the center of the simulation domain. Net $\dot{M}_{\rm disk}$ is calculated as the sum of the star formation rate $\dot{M}_\ast$ and the rate of gas change $\dot{M}_{\rm gas}$. If $\dot{M}_{\rm disk}=0$, gas consumption by star formation precisely accounts for all the gas lost from the disk. If $\dot{M}_{\rm disk}<0$, star formation accounts for only part of the gas loss, meaning that outflows have removed some of the disk gas. And if $\dot{M}_{\rm disk}>0$, then consumption of gas by star formation is more than compensated by CGM accretion and the gas shed by stars, although the latter process contributes on the order of $0.1\ \Msun\ \mathrm{yr^{-1}}$ or less over $t=2$--4~Gyr.

We compare $\dot{M}_{\rm disk}$ to the star formation rate on the right axis, but the SFR has been binned at a higher rate than was used for the net $\dot{M}_{\rm disk}$ calculation ($\Delta t=0.6$~Myr for the SFR instead of $\Delta t=50$~Myr for $\dot{M}_{\rm disk}$). A pink line shows the smoothed SFR that was used in Figure \ref{fig:mass_ev}. Horizontal bars show the average $\dot{M}_{\rm disk}$ over 1~Gyr.

This figure makes it clear that the \textsc{CoolFlow} variant's persistent star formation (as seen in Figure \ref{fig:mass_ev}) is accompanied by a persistent inflow of gas. This is the expected behavior for a simulation without feedback. Intriguingly, the \textsc{NoRot} simulation also sees a net growth in its disk gas from 2--3~Gyr. This corresponds to the large growth (and then steady decline) in gas mass seen in Figure \ref{fig:mass_ev_late}.

The \textsc{HighRatio} variant has the most stable disk mass of the variants, being the simulation with $\dot{M}_{\rm disk}$ that is consistently closest to zero. The other variants exhibit much larger fluctuations (even \textsc{CoolFlow}, though its net $\dot{M}_{\rm disk}$ is almost always positive). Yet for the variants with feedback, averaging the net $\dot{M}_{\rm disk}$ over Gyr timescales brings the net rate of change in disk mass closer to zero.

Visually, there is no clear time correlation between net $\dot{M}_{\rm disk}$ and SFR. Sometimes $\dot{M}_{\rm disk}$ decreases following a burst of star formation, such as the fiducial simulation at $t\approx2.4$ and 3.1~Gyr, \textsc{LinRot} at 2.6~Gyr, \textsc{NoRot} at 2.2~Gyr, and \textsc{HighRatio} at 3.4~Gyr. Other times, a dip in $\dot{M}_{\rm disk}$ \textit{precedes} a burst of star formation, as in \textsc{NoRot} at $t\approx3.1$~Gyr and \textsc{LinRot} at 3.8~Gyr. Bursts of high instantaneous SFR and more prolonged periods of steady star formation both tend to be close in time to large changes in $\dot{M}_{\rm disk}$, but the sign of this change is not consistent, nor is the magnitude or whether the star formation precedes or follows a $\dot{M}_{\rm disk}$ change. This lack of a clear correlation will be discussed more in Section \ref{sec:self-reg}.

In Figure \ref{fig:tcool_dist} we look at the cooling time distribution of CGM gas. We define the CGM as a sphere with radius $r_{200}=206$~kpc, with a cylinder 40~kpc in diameter and $8z_{\rm s} = 2.6$~kpc thick excised to remove the disk. This is consistent with the region used for the disk in Figures \ref{fig:mass_ev}, \ref{fig:mass_ev_late}, and \ref{fig:net_gas_flow}. The cumulative mass distributions are presented in intervals of 1~Gyr, starting from the initial conditions. The stellar feedback efficiency is ramped from 1--2~Gyr as explained in Section \ref{sec:FB-ramp}. The second column therefore represents the cooling time structure of the CGM after it has had a chance to evolve and interact with the disk, but before the impact of feedback. Dashed lines show the cooling time mass distribution when the disk is \textit{not} removed (but we still restrict to gas with $r<r_{200}$).

The \textsc{LowRatio} and \textsc{HighRatio} variants have higher and lower CGM masses than the fiducial simulation, respectively, as noted in Section \ref{sec:variants}. Indeed, the initial $\tctff$ ratio is the only thing that affects the initial cooling time distribution. 

The overall mass of the CGM drops over time in all variants, irrespective of feedback. The majority of the mass loss (all but $\sim2\times10^5\ \Msun$) is an artifact of the initial conditions: as explained in Section \ref{sec:edge-pics}, the collapse of the initial disk pushes back on the CGM, inflating it and pushing gas beyond $r_{200}$. Indeed, by 4~Gyr, the CGM mass within $r_{200}$ has dropped by over an order of magnitude, with the fiducial CGM mass dropping to $\sim1.8\times10^9\ \Msun$.

At 1~Gyr, we can see that each variant has increased its amount of CGM gas with $t_{\rm cool}\sim0.4$~Gyr. Given the difference between the solid and dashed lines at $t_{\rm cool}\lesssim0.4$~Gyr and the flatness of the disk-included profile in this region, most of the very low cooling time gas has been accreted onto the disk. The \textsc{NoRot} simulation is the one variant with notably more gas with $t_{\rm cool}\lesssim0.4$~Gyr left in its CGM at 1~Gyr. This is intriguing given its rapid growth in stellar mass seen in Figure \ref{fig:mass_ev}.

From 2--3~Gyr, the \textsc{CoolFlow} simulation has less CGM gas at short cooling times ($t_{\rm cool}<1$~Gyr) than in the fiducial, but a roughly equivalent amount of gas at longer cooling times. This is the imprint of the \textsc{CoolFlow} simulation's constant inflow of gas (Figure \ref{fig:net_gas_flow}), which depletes gas with short cooling times. The fiducial simulation experiences gas inflow as well of course, but in a reduced capacity thanks to stellar feedback: at the 0 and 1~Gyr snapshots---before the feedback efficiency is increased---the fiducial and \textsc{CoolFlow} variants have essentially identical cooling time-mass distributions. The distributions (both with and without the disk) deviate at 2~Gyr after feedback has become effective. This also means that any variations in cooling time distribution before 2~Gyr are due to CGM differences. Interestingly, the fiducial and \textsc{CoolFlow} CGM distributions come roughly back into agreement at 4~Gyr (though the fiducial simulation has a more massive disk).

Figure \ref{fig:tcool_norm} highlights the differences in shape between the CGM cooling time-mass distributions of our variants by normalizing them. On the left we show the $\tctff$ variants: the fiducial, \textsc{LowRatio}, and \textsc{HighRatio} simulations. On the right are our rotation variants: the fiducial, \textsc{LinRot}, and \textsc{NoRot} runs.

We first focus on the $\tctff$ variants. At 1~Gyr, the \textsc{LowRatio} variant has more of its mass in gas with $1 < t_{\rm cool} <10$~Gyr than the other two, and slightly less in gas with $t_{\rm cool} < 1$~Gyr. The excess of gas with $t_{\rm cool}\sim1$~Gyr seems consistent with the large, positive $\dot{M}_{\rm disk}$ exhibited by the \textsc{LowRatio} simulations from $t=2$--2.5~Gyr in Figure \ref{fig:net_gas_flow}. The fiducial simulation also has positive $\dot{M}_{\rm disk}$ during this time, though of less magnitude than \textsc{LowRatio}. Furthermore, the fiducial simulation has an intermediate amount of CGM gas with $1 < t_{\rm cool} <10$~Gyr. These differences in shape are accompanied by differences in absolute mass of gas with $1 < t_{\rm cool} <10$~Gyr. It nevertheless seems an intuitive result that the (relative and absolute) amount of gas with $t_{\rm cool}$ of a few gigayears affects the amount of gas accreted by a disk a few gigayears in the future.

By 3~Gyr, we see that the three simulations have evolved cooling time-mass distributions with the same general shape. These similarities are for the most part retained by 4~Gyr. This suggests that, despite initial differences in the CGM's structure, feedback has impacted their cooling-time mass distributions in a similar manner. This is despite the morphological differences in their feedback (Figure \ref{fig:edge_comp}). 

The similarity between the fiducial and \textsc{CoolFlow} simulations in Figure \ref{fig:tcool_dist} tells us that the \textsc{LowRatio}, \textsc{HighRatio}, and fiducial cooling time distributions have the same shape because \textit{feedback does not affect the cooling time mass distribution much at all.} This seems strange given the dramatic outflows in Figure \ref{fig:edge_comp}, but most of that outflowing gas has $t_{\rm cool}$ longer than a Hubble time and therefore should not be expected to return to the disk. The gas that is visible in Figure \ref{fig:tcool_dist}, particularly that with $t_{\rm cool}\lesssim10$~Gyr, is gas that has necessarily not been heated by outflows. This is likely gas near the disk midplane which avoids being heated.

What \textit{does} have an impact on the CGM's cooling-time structure is rotation, as seen in the right side of Figure \ref{fig:tcool_norm}. Not only is there a difference between \textsc{NoRot} and the other variants with rotating CGMs, but there are also variations due to the different rotation profiles. We adopt a very straightforward prescription for the CGM's rotation (Equation \ref{eq:rot}), but these differences highlight the importance of understanding the angular momentum of the CGM.

\section{The Circumgalactic Medium}\label{sec:cgm}

The CGM has a complex, multiphase structure that can be difficult to encapsulate. The digestability of information must be balanced against the loss of detail. We can see in Figures \ref{fig:fid_ev} and \ref{fig:edge_comp} that our simulations do not have a spherically symmetric CGM and are instead closer to having cylindrical symmetry. Trying to encapsulate the structure of the CGM with, e.g., mass-weighted spherical profiles therefore constitutes a loss of information. Furthermore, averages are biased towards the highest values, even when mass-weighted.

\begin{figure}[tp]
    \centering
    \includegraphics[width=\columnwidth]{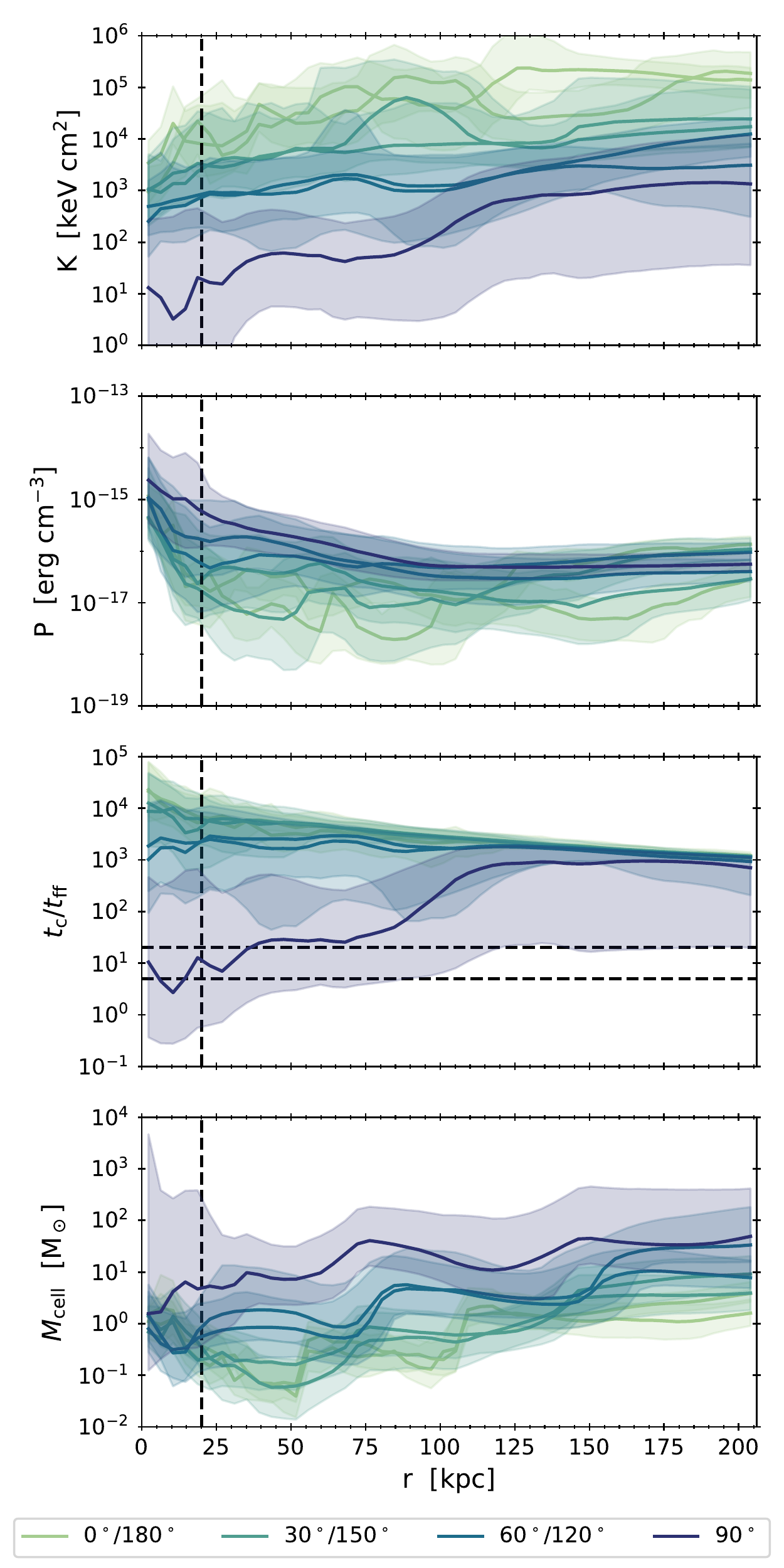}
    \caption{Entropy, pressure, $\tctff$, and gas mass radial profiles of the fiducial CGM as a function of polar angle $\theta_{\rm cen}$ at late times ($t=3$--4~Gyr). The disk midplane corresponds to $\theta=90^\circ$. Solid lines are the time average of the 50th percentiles (medians), and shaded regions are of the 16th and 80th percentiles; see text for more details on our analysis. The vertical dashed line at 20~kpc represents the approximate extent of the disk adopted throughout this text, while the horizontal dashed lines show the $\tctff$ range of 5--20 predicted by precipitation theory.
    }
    \label{fig:all_theta}
\end{figure}

Therefore, in an attempt to encapsulate the radial structure of the CGM, we subdivide the gas within the virial radius based on its polar angle $\theta$. This creates a set of cones (at the poles) and circular wedges. We choose to have seven regions subtending $\Delta \theta =30^\circ$; for example, the region with $\theta\in(0^\circ,15^\circ)$ covers the top pole of the simulation domain, and $\theta\in(75^\circ,105^\circ)$ encompasses the disk midplane. To keep our notation concise, we will refer to these regions based on their central polar angle, $\theta_{\rm cen}$. For the previous examples, these would be $\theta_{\rm cen}=0^\circ$ and $\theta_{\rm cen}=90^\circ$.

\begin{figure}
    \centering
    \includegraphics[width=\columnwidth]{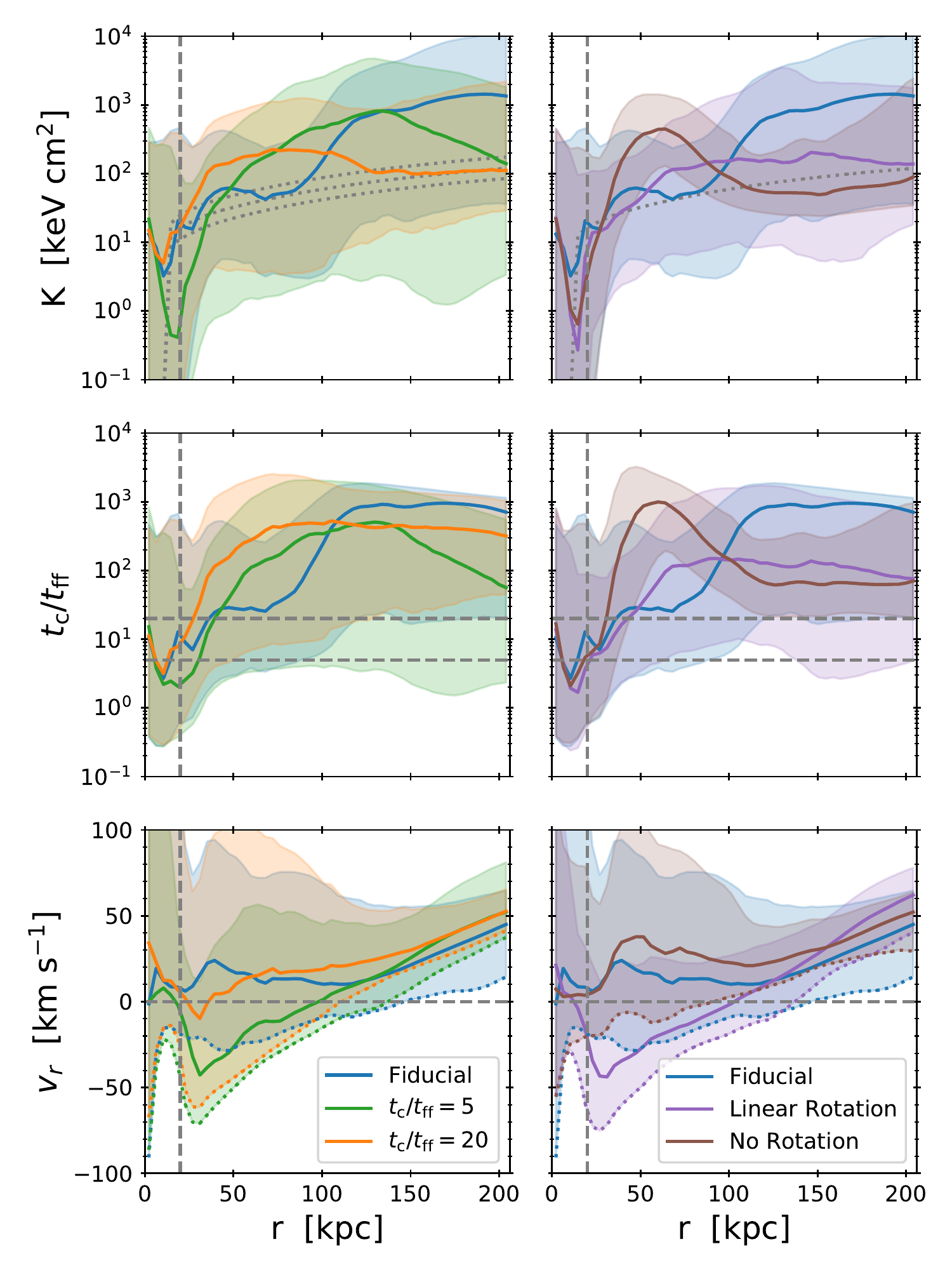}
    \caption{Entropy, $\tctff$, and radial velocity profiles for the cones at $\theta_{\rm cen}=90^\circ$ (i.e., covering the disk midplane), averaged over $t=3$--4~Gyr, for the simulation variants with stellar feedback. Solid and shaded lines are the same as in Figure \ref{fig:all_theta}. The vertical dashed line at 20~kpc marks the approximate edge of the disk, and the horizontal dashed lines mark $\tctff=5$--20 and $v_r=0$~km/s. The initial entropy profiles are shown as dotted gray lines, and dotted colored lines highlight the time-averaged 16th percentiles in radial velocity.
    }
    \label{fig:theta_90}
\end{figure}

Within each region, we radially bin quantities of interest from $r=2$ to $r=206 \ \mathrm{kpc}\, \approx r_{200}$ using 51 bins. This gives a bin width of $\Delta r\approx4$~kpc. For each radial bin, we find the 16th, 50th, and 84th percentiles. This analysis is done for each simulation output (whose cadence is $\Delta t=50$~Myr). Figures \ref{fig:all_theta} and \ref{fig:theta_90} show the time-average of these percentiles from $t=3$ to $t=4$~Gyr. We restrict the average to the last gigayear as the feedback ramp is at its full strength and the profiles are visually the most stable over time.

Figure \ref{fig:all_theta} shows the radial profiles of entropy, pressure, $\tctff$, and gas mass for the fiducial simulation. A vertical dashed line at $r=20$~kpc marks the rough edge of the disk used throughout our analysis. The horizontal dashed lines in the $\tctff$ panel indicate the 5--20 range predicted by precipitation \citep{voit17-GlobalModelCircumgalactic, voit18-RoleTurbulenceCircumgalactic,voit21-GraphicalInterpretationCircumgalactic}. For $\theta_{\rm cen}=90^\circ$ (the disk midplane), the 16th percentile of entropy is cut off at $r\sim 30$~kpc, reaching values of $\sim 10^{-4}$~keV~cm$^{-2}$ at the smallest radii. This is the influence of the galactic disk, which, unlike in much of our analysis, we do \textit{not} excise. The disk is also evident in the 84th percentile of the mass profile.

It is clearly visible that all four quantities vary with polar angle: entropy and $\tctff$ decrease towards the disk midplane at $90^\circ$ while cell mass and pressure rise. The simulation is roughly symmetric about the disk plane. The higher entropy, lower pressure, and lower mass near the poles are consistent with the bipolar outflows seen in Figure \ref{fig:edge_comp}, while the opposite trends at the disk midplane are consistent with the inflow seen along the midplane. All simulation variants demonstrate these trends. Near the poles (far from the disk midplane), the time-averaged $\tctff$ ratio tends toward a smoothly declining profile. At these high angles, the time-averaged cooling time is approximately constant at $t_{\rm c} \sim 2\times 10^3$~Gyr. The shape of $\tctff$ is set primarily by the freefall time and, by extension, the NFW halo profile. On the other hand, only the $\tctff$ profile centered on $90^\circ$ (the plane of the disk) is remotely consistent with the $\tctff$ range of 5--20 predicted by precipitation. This is also the polar angle with the highest cell mass, and is the angle at which most of the cold gas inflow is located.

The mass dominance of the $\theta_{\rm cen}=90^\circ$ profiles, as well as their association with the inflowing gas along the disk midplane in Figure \ref{fig:edge_comp}, motivates isolating these profiles for comparison across the simulation variants. This is done in Figure \ref{fig:theta_90}, where the time-averaged profiles of entropy, $\tctff$, and radial velocity for the $\theta \in (75^\circ,105^\circ)$ wedge are shown for all of the simulation variants with stellar feedback. The time averaging is again restricted to the last 1~Gyr of simulation evolution. A vertical line at 20~kpc again marks the nominal edge of the disk, and horizontal dashed lines mark $\tctff=5$--20 and $v_r=0$~km/s. The initial entropy profiles are shown as dotted gray curves. Colored dotted lines are used to highlight the 16th percentile of the radial velocity, which is predominantly negative. The entropy profile within the disk is once again cut off to better demonstrate the CGM. 

For all simulations, the spread in entropy and $\tctff$ between the time-averaged 16th and 84th percentiles crosses several orders of magnitude. Though this spread is still less than the overall dependence on polar angle (Figure \ref{fig:all_theta}), we can see from the radial velocity that our wedges encompass both inflowing and outflowing gas. Figures \ref{fig:fid_ev} and \ref{fig:edge_comp} show that the inflow region is very thin. We chose a large $\Delta \theta$ in order to capture potential warping of this region (most evident in the fiducial simulation; see the 3rd and 4th columns of Figure \ref{fig:fid_ev}) and so inevitably we capture a mix of gas phases.

Focusing on the $\tctff$ ratio, we see that the \textit{median} of gas near the midplane is predominantly higher than predicted from precipitation theory, though the 16th percentile does extend into the 5--20 range for all but the \textsc{NoRot} simulation. Precipitation requires a local $\tctff\sim1$ for gas to actually condense, which may not be captured by the 16th percentile. Yet this Figure is strong evidence that our simulated galaxies are \textit{not} being regulated by precipitation, even if we search for precipitation \textit{only} in the disk midplane.

\section{Discussion}\label{sec:discussion}

We simulated a suite of idealized galaxies that are similar to the Milky Way in order to explore the conditions under which self-regulating feedback might arise. Our hypothesis was that these galaxies would naturally regulate their star formation rates according to the predictions of precipitation theory if initialized with a CGM having $\tctff \sim 10$ \citep{voit19-AmbientColumnDensities}. This work therefore complements that of \citet{prasad20-EnvironmentalDependenceSelfregulating}, who studied the precipitation-regulation of larger AGN-dominated systems with SNIa feedback (i.e., massive central ellipitcal galaxies in groups and clusters), in that it explores the ability of less massive galaxies to regulate themselves through stellar feedback alone.

In light of the results from \citet{prasad20-EnvironmentalDependenceSelfregulating}, the question motivating our work is: can we also create a self-regulating galaxy-CGM system but with feedback coupled to star formation instead of AGN and older stellar populations?  We broadly take self-regulation to mean that feedback tunes the net inflow of CGM gas to match the disk's time-averaged star formation rate. Since our only feedback mechanism is Type II SNe, this also implies that the star formation rate of self-regulated systems would be tuned to the inflow of cold gas. 

Ultimately, we do not consider the galaxies we have simulated to be self-regulating. Instead, the CGM in our simulations experiences large-scale disruption due to outflows. Only gas along the disk midplane is of low enough entropy to be able to cool and accrete onto the disk. Even this limited accretion mode does not appear to be precipitation-regulated based on the most general measure, the median ratio of cooling and freefall times. Our star formation rates drop to very low values of order $0.1\ \Msun$/yr after the onset of our feedback efficiency ramp. This limited gas accretion and star formation keep the average disk growth, $\dot{M}_{\rm disk}$, near zero. Rather than creating a system that maintains a moderate SFR, we have created a system in which feedback essentially shuts off star formation. 

We start our discussion with Section \ref{sec:cgm-impact}, where we examine the impacts of variations in the simulation parameters. Then, in Section \ref{sec:compare}, we compare the structural features of our simulations to other works, both observational and theoretical. Section \ref{sec:toomre} highlights the role the Toomre instability plays in the evolution of our galaxies. Our simulations do \textit{not} have self-regulated star formation, but their failure to self-regulate is illuminating. In Section \ref{sec:self-reg} we interrogate the definition of self-regulation and highlight the ambiguity as to the expected timescales involved. Finally, in Section \ref{sec:missing-elements} we discuss the physical effects missing in our simulations that could better couple gas accretion and stellar feedback.

\subsection{Impact of CGM Variations} \label{sec:cgm-impact}

The overall behavior of our simulations is unaffected by variations in the CGM: once stellar feedback is at its full strength the CGM becomes disrupted by wind-driven bubbles and the SFR drops by an order of magnitude. The exception is, of course, the \textsc{CoolFlow} variant which completely lacks stellar feedback. Yet despite observing no impact on the overall behavior, we do see important differences manifest in our simulation variations. The variations in initial conditions exert their biggest influence at $t<1$~Gyr, when feedback is made artificially weak. In the following subsections we will discuss the differences resulting from changes to the initial $\tctff$ ratio (Section \ref{sec:var_tctff}) and from alterations to the CGM's initial rotation profile (Section \ref{sec:var_rot}).

\subsubsection{Variation in Initial $\tctff$}\label{sec:var_tctff}

The \textsc{LowRatio} and \textsc{HighRatio} variants are distinguished by modifying the condensation criterion $\tau$ in Equation \ref{eq:disk}. This parameter is essentially the initial $\tctff$ ratio of the CGM, though as discussed in Section \ref{sec:CGM-ICs} the actual $\tctff$ ratio deviates slightly. Including the fiducial simulation, we sample $\tau \in [5,10,20]$.

Because the CGM's density structure is set by $\tau$, these variants have different starting CGM masses. The \textsc{LowRatio} variant has the most mass at $3.9\times10^{10}\ \Msun$, followed by the fiducial with $2.3\times10^{10}\ \Msun$ and the \textsc{HighRatio} with $1.3\times10^{10}\ \Msun$.\footnote{
The CGM is defined as a sphere with $r=r_{200}\approx206$~kpc with a cylinder excised for the disk. This disk has height $z=4z_{\rm s}=1.3$~kpc from the midplane and radius $R=[27.5,28.5,29]$~kpc for the \textsc{LowRatio}, fiducial, and \textsc{HighRatio} variants respectively.
} Therefore, a higher $\tau$ results in an overall less massive CGM. 

Moreover, a higher $\tau$ lowers the relative amount of gas with initial $t_{\rm cool} < 1$~Gyr. This is seen in Figure \ref{fig:tcool_dist}. For a real multiphase CGM, this can be understood via the framework of \citet[]{voit21-GraphicalInterpretationCircumgalactic}: the median $\tctff$ of the CGM defines the center of a \textit{distribution} in $\tctff$. Moving the median to higher values means that less of the distribution covers low cooling times, and therefore less gas is able to efficiently cool, condense, and reach the galaxy. Our simulations, however, do \textit{not} start with a multiphase CGM; rather, the temperature and density are smooth, spherically symmetric functions of radius. In fact, the initial temperature profile is very similar between the fiducial, \textsc{LowRatio}, and \textsc{HighRatio} simulations. The differing masses of gas with $t_{\rm cool} < 1$~Gyr are then a result of both variations in the initial density profiles and the different initial $\tctff$. 

The initial stellar mass growth of the \textsc{HighRatio} and \textsc{LowRatio} simulations is easily described by the different masses of gas with $t_{\rm cool} < 1$~Gyr. This is seen in Figure \ref{fig:mass_ev}, where at $t<1$~Gyr the \textsc{LowRatio} simulation has the highest disk gas mass, greatest growth in stellar mass, and highest SFR. Conversely, the \textsc{HighRatio} variant has the least growth in stellar mass and lowest SFR, although its gas mass is not as distinct from the fiducial simulation as the \textsc{LowRatio} simulation is.

The difference in low cooling time gas also seems to have a slight impact on the physical size of the disks, as seen in Figure \ref{fig:face_comp} and Table \ref{tab:SFradii}. The \textsc{LowRatio} simulation has the most gas with $t_{\rm cool} < 1$~Gyr and correspondingly has the largest radius within which star formation occurs. The converse is not quite true for the \textsc{HighRatio} simulation, whose average radius of star formation is not statistically significantly different from the fiducial simulation's. The average star formation radii in Table \ref{tab:SFradii} are determined over $t=2$--4~Gyr, after the feedback ramp has ended and the bulk of star formation has occurred. This therefore suggests that the initial differences in CGM gas accretion between the \textsc{LowRatio}, \textsc{HighRatio}, and fiducial simulations have lingering effects on the structure of the galactic disk. This effect would appear to be on the same scale as natural variation in the maximum radius of star formation. These differences are likely not a result of the difference in initial disk radius, as the initial \textsc{LowRatio} disk is about 1~kpc \textit{smaller} than the fiducial's gas disk.

Feedback does not generally seem to have much impact on the $t_{\rm cool}$ distribution function for CGM gas. This is seen between the $\tctff$ variants in Figure \ref{fig:tcool_norm}. Though the $\tctff$ variants have different star formation histories and stellar masses, feedback neither amplifies nor diminishes any pre-existing differences in the $t_{\rm cool}$ distribution function.
The similarity between the fiducial and \textsc{CoolFlow} distributions at $t=4$~Gyr further indicates that feedback has minimal impact on the CGM's overall availability of low cooling time gas.

Even though the cooling time distribution is largely unaffected by feedback, the $\tctff$ variants do not evolve identically. In Figure \ref{fig:edge_comp} we can see that outflows in the \textsc{LowRatio} simulation have traveled less far than in either the fiducial or \textsc{HighRatio} simulations. For $t<1$~Gyr, the \textsc{LowRatio} simulation has the highest CGM pressure of the $\tctff$ variants, and \textsc{HighRatio} the lowest. This generally tends to remain true for the outer CGM ($r\approx150$--200~kpc) even as outflows drop the overall CGM pressure. 

The $\dot{M}_{\rm disk}$ in Figure \ref{fig:net_gas_flow} also shows interesting differences. The \textsc{HighRatio} variant experiences the smallest $\dot{M}_{\rm disk}$ fluctuations of the simulations considered here, suggesting that relatively little mass is involved in the cycle of accretion and star formation. The fluctuations of $\dot{M}_{\rm disk}$ in the \textsc{LowRatio} simulation are on par with and occasionally larger than those experienced by the fiducial simulation. Generally, the fluctuations in $\dot{M}_{\rm disk}$ get smaller as the simulations go on, as seen by the average from $t=3$--4~Gyr. 

It may be that the \textsc{LowRatio} simulation, being able to accrete more cool gas initially, was set onto a cycle of large amounts of gas accretion followed by productive periods of star formation, rather like an oscillator with a large initial perturbation. Also like most oscillators found in nature, the \textsc{LowRatio} simulation experiences damping in its cycle of accretion and star formation. Though it is the simulation with the highest stellar mass, Figure \ref{fig:mass_ev_late} shows that it had only moderate gains in stellar mass after 2~Gyr. This figure also shows that the \textsc{LowRatio} simulation continued to accrete gas from $t=2$--3~Gyr, despite the stronger feedback. This may be the \textsc{LowRatio} simulation ``refueling'' in order to maintain a more moderate SFR at the end of the simulation. On the other hand, the fiducial simulation gains the most stellar mass over $t=2$--4~Gyr in Figure \ref{fig:mass_ev_late}, while the \textsc{HighRatio} simulation is with \textsc{LowRatio} near the bottom of the pack, although it is something of a ``late bloomer.'' It may be that the fiducial simulation has the most ideal conditions of these variants for sustained star formation growth. Whether or not we consider it ``self regulating'' is a question we defer to Section \ref{sec:self-reg}.

To summarize, the \textsc{LowRatio}, \textsc{HighRatio}, and fiducial simulations all exhibit a number of differences that are a consequence of how the condensation criterion $\tau$ 
determines both the initial $\tctff$ ratio and density structure of the CGM. These density differences result in differences in total mass. The \textsc{LowRatio} simulation not only starts with more gas, but a larger fraction of it is able to cool efficiently. The converse is true for the \textsc{HighRatio} simulation. These differences are persistent even in the face of feedback. The structure of the disk and its star formation therefore retain their early differences throughout the simulation runtime.

\subsubsection{Variation in CGM Rotation}\label{sec:var_rot}

We now consider variations that arise from the CGM's rotation. Apparent from Figure \ref{fig:mass_ev} is that both the presence of rotation, as well as its variation with radius, has an impact on the ability of the CGM to supply gas to the disk.

We have indicated throughout this work that the \textsc{NoRot} simulation should be able to accrete gas more efficiently because its CGM gas does not have to shed angular momentum to reach the disk. This is evident in the stellar growth of Figure \ref{fig:mass_ev} at $t<1$~Gyr. It may also explain the large growth in gas mass seen over $t\sim2$--3~Gyr in Figures \ref{fig:mass_ev_late} and \ref{fig:net_gas_flow}. 
In Section \ref{sec:edge-pics}, we noted a remnant of the initial disk conditions that is present in the CGM: due to insufficient vertical support against gravity, the initial disk collapses and spreads outwards into the CGM. The \textsc{NoRot} variant's initial disk has the smallest amount of radial spread due to the lack of rotation in its CGM. 

The mere inclusion of rotation in the CGM has a large effect on gas accretion and simulation evolution, but the radial profile of that rotation also has an impact. The \textsc{LinRot} simulation forms more stars than the fiducial model within the first 1~Gyr, before stellar feedback has much effect. The \textsc{LinRot} simulation also has lower angular momentum at all CGM radii, making it easier for gas to accrete than in the fiducial simulation. Rotation is also able to change the shape of the cooling time-mass distribution in Figure \ref{fig:tcool_norm}, affecting the availability of gas that can be accreted in addition to the ease of accretion.

\citet[]{su20-CosmicRaysTurbulence} also performed isolated galaxy simulations that included a rotating CGM. These simulations targeted cool-core clusters with halo masses between $10^{12}$ and $10^{14}\  \Msun$. Rotation followed a $\beta$-profile with $\beta=1/2$, set to be twice the net dark matter spin. Rotation therefore comprised 10-15\% of the CGM's support against gravity, with the rest being supplied by thermal energy. This is in contrast to our simulations, where rotation was not considered in the calculation of hydrostatic balance. The simulations of \citet[]{su20-CosmicRaysTurbulence} were included in the \citet[]{fielding20-FirstResultsSMAUG} meta analysis, where it is noted that these simulations had an enhanced cold phase at $r < 0.2 r_{200}$ compared to the other isolated galaxy simulations in the analysis. This enhancement is attributed to CGM rotation---which was not included in \citet[]{fielding17-ImpactStarFormation} and \citet[]{li20-HowSupernovaeImpact}---in agreement with our results.

\citet[]{defelippis20-AngularMomentumCircumgalactic} studied the CGM angular momentum of Milky Way-mass galaxies in IllustrisTNG. They split these galaxies into samples based on the stellar specific angular momentum, resulting in high- and low-momentum populations. Both populations have inflowing cold CGM gas near the disk plane. In the sample with high specific angular momentum, this inflow is well aligned with the disk; however, the radial inflow disappears for $r<0.2 r_{200}$. Much like with our simulations, it is suggested that this is due to the presence of strong rotation. We note that our rotation speeds \citep[derived from][]{hodges-kluck16-RotationHotGas} are higher than the rotational speeds measured by \citet[]{defelippis20-AngularMomentumCircumgalactic}.

\subsection{Comparison of Structural Features to Other Works}\label{sec:compare}

It is worth comparing some of the structures seen in our simulations with those seen or predicted in other works. These works include observations, analytic predictions, and cosmological simulations. We'll begin with the gas disk, expand outward to the accretion flows seen along the disk midplane, and finally consider the structure of our feedback-driven outflows.

\subsubsection{Extended Gas Disk}\label{sec:extended_disk}

Starting with the disk, we can see in Figure \ref{fig:face_comp} that there is cold, dense, rotating gas that extends beyond the average radius of star formation in all of our variants. This is consistent with observations of spiral galaxies in the THINGS survey \citep{leroy08-StarFormationEfficiency}. These observations show that spiral galaxies have high star formation efficiency in their H$_2$-dominated cores. This efficiency declines with radius in the extended neutral hydrogen disk. 

The structure of our galactic disk is also consistent with the findings of \citet[]{lopez20-SlicingCoolCircumgalactic} and \citet[]{tejos21-TelltaleSignsMetal}. These papers present two instances of extended Mg II disks ($r\sim20$--40~kpc) that are co-rotating with the interstellar medium. Our extended cool disks are of order $r\sim20$-30~kpc. We note that, while an extended cool disk is an emergent feature of our simulations, we purposely aligned the angular momentum vectors of the CGM and the disk. It is unclear what impact misaligned angular momentum vectors might have on our extended gas disk.

\subsubsection{Accretion Along the Midplane}\label{sec:planar}

Figures \ref{fig:fid_ev} and \ref{fig:edge_comp} make it clear that, while inflowing gas can be found throughout the CGM, only inflowing gas near the disk midplane is able to reach the disk. Inflowing gas \textit{not} along the midplane is disrupted by outflows. The exception to this is of course the \textsc{CoolFlow} variant, which lacks stellar feedback. 

Inflow within the disk plane is not unprecedented in cosmological simulations. \citet[]{trapp22-GasInfallRadial} find this inflow mode in the FIRE-2 simulations and note that it is the dominant source of accretion there just as it is in our simulations. Gas accretion in the EAGLE simulations is also anisotropic, favoring low heights relative to the disk and inflow speeds of 20--60~km/s \citep[][]{ho19-HowGasAccretion}. This inflow is predominantly cold gas found near the disk or in low-angular momentum streams, and estimates of the mass rate are enough to meet or exceed the SFR.

Within IllustrisTNG, \citet[]{defelippis20-AngularMomentumCircumgalactic} also see this inflow structure in $10^{11.75}$--$10^{12.25} \Msun$ halos. \citet[]{truong21-PredictionsAnisotropicXray} see a global anisotropy in the CGM, with density enhanced parallel to the disk plane, and temperature and metallicity enhanced along the orthogonal minor axis. Their anisotropies are more subtle than what we see in our own simulations, being of only 0.1--0.3~dex. These anisotropies peak in Milky Way-like galaxies with $M_\ast \sim 6\times10^{10}\ \Msun$, which is the mass at which super massive black hole feedback turns on in TNG. Generally, the anisotropies are larger in galaxies with SMBH feedback, except for metallicity: the metallicity anisotropy is more pronounced in star-forming and disky galaxies.

The inflowing gas spans a temperature range from $\sim5\times10^3$--$5\times10^5$~K, with the coldest gas living in denser filaments. Gas condenses and cools while being part of a rotating inflow. This is reminiscent of the accretion mode observed in \citet[]{hafen22-HotmodeAccretionPhysics} and described in detail in \citet[]{stern21-VirializationInnerCGM}. Gas inflow is hot until it reaches $\sim20$~kpc scales, at which point angular momentum slows its inward motion. Radiative cooling then exceeds heating due to compression, and the gas temperature drops to $\sim10^4$~K or below. Importantly, \citet[]{hafen22-HotmodeAccretionPhysics} note that the angular momentum of inflowing gas aligns itself with the galaxy before cooling, but because the angular momentum of our CGM and disk are constructed to be aligned, we cannot make any comparison on this point.

\subsubsection{Outflow Structure}

Our simulations are dominated by outflows to an unrealistic degree. Very low density cavities extend out to the virial radius by $t=4$~Gyr. Outflows have a very wide opening angle of essentially $180^\circ$, covering the face of the disk. The CGM essentially becomes obliterated above and below the disk.

That said, the structure of our outflows matches well to analytic models from \citet[]{lochhaas18-FastWindsDrive}. These models describe how relatively slow-moving gas bubbles can be inflated by fast galactic winds. In their model, winds drive a forward shock through the CGM. Their interaction leads to the development of a reverse shock, and between these two fronts lies a contact discontinuity. This discontinuity separates material driven out by the winds from the swept-up CGM gas. \citet{lochhaas18-FastWindsDrive} refer to the region between the reverse shock and the contact discontinuity as the ``shocked wind.'' It is this shocked wind that is responsible for the growth of the gas bubble. Immediately behind this shocked wind (inside the reverse shock) lies cool, unshocked gas, followed by hot unshocked gas immediately next to the galaxy.

We can clearly see the shocked wind in the density projections of Figure \ref{fig:fid_ev}, often with winds inside winds. The models of \citet{lochhaas18-FastWindsDrive} use a continuous wind, while our simulations have multiple discrete winds driven by episodic Type II supernova feedback. At 2~Gyr, a band of shocked winds is visible both above and below the disk. Subsequent winds are able to travel faster due to the evacuated region that follows behind the first \citep{lochhaas18-FastWindsDrive}. The thickness of the outermost shocked wind grows as its reverse shock propagates. It travels with speeds on the order of $\sim100$~km/s, which is an order of magnitude lower than the feedback-driven winds, corroborating the major result of \citet{lochhaas18-FastWindsDrive}. The density of the shocked wind is also relatively constant across its width as assumed by the analytic models, though that density drops as the shocked wind expands. 

Instead of a layer of cold unshocked gas, the gas behind the shocked wind is very hot ($\gtrsim 5\times10^6$~K) and chaotic thanks to successive feedback events. Additionally,  these continued outflows prevent the outermost shocked wind from stalling and falling back onto the galaxy, as seen in the additional tests run for Section 5.3 of \citet{lochhaas18-FastWindsDrive}. They also appear to disrupt any dense, cooling material that may fall inward towards the galaxy. \citet[]{lochhaas18-FastWindsDrive} note that such gas could be seeded by a Rayleigh-Taylor or Vishniac \citep{vishniac83-DynamicGravitationalInstabilities,vishniac89-StabilityDeceleratingShocks} thin-shell instability. While we see plenty of turbulence, we see no evidence of infalling material reaching the galactic disk outside of the disk midplane.

\subsection{The Role of the Toomre Criterion}\label{sec:toomre}

\begin{figure}
    \centering
    \includegraphics[width=\columnwidth]{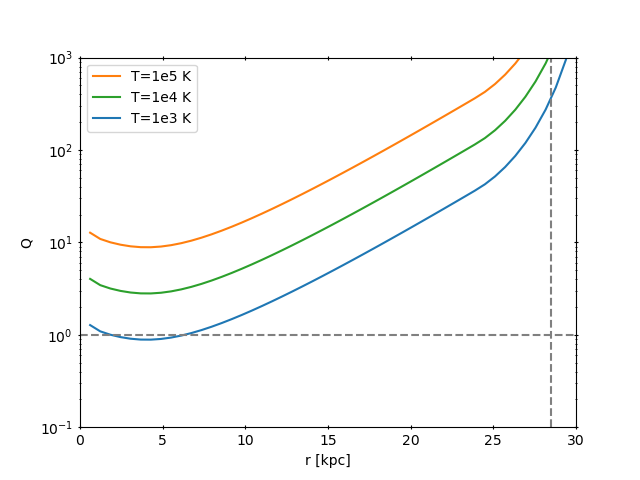}
    \caption{Toomre parameter $Q$ as a function of radius for our simulation initial conditions. We calculate $Q$ analytically based on the static background gravitational potentials and the initial disk density profile (Section \ref{sec:disk-ICs}). The disk is initially isothermal with $T=10^5$~K (orange), but other profiles are shown assuming the disk cools uniformly to $10^4$~K (green) and $10^3$~K (blue). The latter is just below the minimum temperature for star particles to form ($3\times10^3$~K). The vertical dashed line marks the edge of the fiducial disk, based on how it blends with the CGM.}
    \label{fig:toomre}
\end{figure}

One of the most striking features of Figure \ref{fig:mass_ev} is that the disk gas mass is almost identical between our variants with feedback. This is true at all times, both during the $t<1$~Gyr star formation burst and late into the simulations, despite vast differences in stellar mass. This feature is consistent with the Toomre criterion $Q$ for disk fragmentation
\begin{equation}\label{eq:toomre}
    Q = \frac{c_s \kappa}{\pi G \Sigma_{\rm g}}
\end{equation}
where $c_s$ is the sound speed, $\kappa$ is the epicyclic frequency of the rotating disk, and $\Sigma_g$ is the gas surface density of the disk's face \citep{toomre64-GravitationalStabilityDisk}. If $Q>1$ at a given disk radius, the gas is stable against fragmentation. If $Q$ locally drops below 1, then at that radius, the disk is able to fragment into clumps that lead to star formation. 

Important for our discussion is the dependence of $Q$ on the gas surface density, $\Sigma_{\rm g}$. As gas accretes, the surface density rises and thus $Q$ drops. Star formation lowers $\Sigma_{\rm g}$, bringing $Q$ back towards 1 and the disk back towards stability. Gas accretion and fragmentation therefore conspire to keep $Q\sim1$, which results in a fairly stable gas surface density and, by extension, overall mass. All disks are constructed to have the same initial $\Sigma_g$ profile through Equations \ref{eq:disk} and \ref{eq:disk_mod} (though the \textsc{HighRatio} and \textsc{LowRatio} disks extend to larger or smaller radii, respectively). They also all evolve within the same static gravitational potential, which sets $\kappa$, and remain roughly isothermal with $T\sim4\times10^3$~K (although $T\sim10^2$~K in the densest clumps), setting $c_s$. The similarity in disk temperature and sound speed between all of the variants indicates that the Toomre criterion is involved in keeping all variants near the same disk mass. The exception to this is the \textsc{CoolFlow} variant, whose disk mass notably deviates from the other variants despite having only a slightly warmer disk overall ($T\sim5\times10^3$~K). The lower disk gas mass in the \textsc{CoolFlow} variant may occur because there is no supernova feedback to generate turbulence, which would raise the \textsc{effective} value of $c_s$ in Equation \ref{eq:toomre}.

Figure \ref{fig:toomre} shows the $Q$ parameter calculated as a function of radius for the simulation initial conditions (Section \ref{sec:disk-ICs}). We perform this calculation analytically (i.e., we do not estimate $Q$ from the simulation itself) using the background stellar and dark matter potentials and the initial disk density profile (Equations \ref{eq:disk} and \ref{eq:disk_mod}) to calculate the epicyclic frequency $\kappa$ and surface density $\Sigma_{\rm g}$, respectively. Since the initial disk is isothermal at $10^5$~K, the sound speed $c_s$ is constant. We also include $Q$ calculated for lower temperatures, assuming the disk cools uniformly: $10^4$ and $10^3$~K. The latter is close to the minimum temperature required for stars to form ($3\times10^3$~K) and is therefore an approximate representation of $Q(r)$ when stars begin to form. We only have $Q\sim1$ within the inner $\sim10$~kpc, which is commensurate with the average star formation radii in Table \ref{tab:SFradii} (with exception of the \textsc{CoolFlow} variant).

\subsection{The Timescale of Self-Regulation}\label{sec:self-reg}

Broadly speaking, self-regulation is a balance between gas inflow and feedback. This definition is used in, e.g., \citet[]{mitchell21-HowGasFlows} and \citet[]{prasad20-EnvironmentalDependenceSelfregulating}. Self-regulation can be understood through analogy to a thermostat, which works to keep the temperature of a room at a specified ``set point.'' Self-regulation in a galaxy should work to keep the star formation rate near a particular set point. What this set point is, exactly, is not set externally (as with a thermostat), but would be set by the conditions of the galaxy and its environment. Thermostats turn off and on a furnace, which in our analogy, is the star formation. The product of a furnace is heat, and the product of our galactic ``furnace'' is stellar feedback.

Real thermostats do not keep a room at the set point all the time. When the temperature drops, the furnace is turned on. When the room gets sufficiently warm (either reaching the set point exactly or maybe slightly above to compensate for cooling), the furnace is shut off. We expect the same thing of galaxies: gas accretion should lead to star formation. The feedback from this star formation will influence the galaxy's ability to accrete gas, modulating future star formation.

The connection between gas accretion and star formation is not temporally or spatially immediate, however. As we discussed in Section \ref{sec:planar}, gas is often seen to accrete along the plane of the disk, but star formation is often concentrated at the center of the disk where densities are highest. Recent estimates put the inward radial mass flow rate at less than 1~$\Msun/\mathrm{yr}$ \citep[][]{diteodoro21-RadialMotionsRadial}. During a feedback burst (when the furnace is on), it is therefore probably not the case that inflow is perfectly balanced by star formation. Rather, the two balance each other \textit{on average over some timescale}.

With Figure \ref{fig:net_gas_flow}, we attempted to identify by eye correlations between the SFR and the net rate of change in disk mass, $\dot{M}_{\rm disk}$, but ultimately found no consistent relation. While there are certainly rigorous statistical tools for addressing time correlations, this analysis is sufficient to raise questions about what precisely one should be looking for in order to decide if a galaxy is self-regulating its star formation. 

The ambiguity over timescale is exacerbated when examining self-regulation with simulations. Because we are limited in resolution and computational power, simulations of Milky Way-mass galaxies must use particles to represent entire stellar populations rather than individual stars. This modeling limitation introduces shot noise into the SFR. It is therefore also worth considering the timescales over which we are concerned with changes in the SFR.

The timescale uncertainties are in large part due to the temporal and spatial separation between gas accretion and feedback. Within our simulations, CGM gas that accretes onto the galaxy is more likely to contribute to a star particle if it reaches the denser disk center. The formation of a star particle  obfuscates the physical and temporal scales involved in the actual formation of molecular clouds and protostars, but these are assumed to be below the temporal resolution of the simulation. Connections between inflow and star formation in simulations seem clearer when made on smaller physical scales, such as gas inflow along dust lanes at the very nucleus of a barred spiral galaxy \citep[][]{moon22-EffectsVaryingMass}. 

Our efforts to directly explore the self-regulation of galaxy star formation has highlighted the need to refine our questions. Specifically, if we as a community are to define self-regulation as a balance between gas accretion and star formation, we should consider on what timescale we expect this balance to be achieved. This way, we may better identify which physical processes contribute to the self-regulation of galaxies.

\subsection{Missing Model Components} \label{sec:missing-elements}

One of the key takeaways from \citet[]{prasad20-EnvironmentalDependenceSelfregulating} and \citet{voit20-BlackHoleFeedback} is that feedback from Type Ia supernovae assists an AGN in self-regulating its own feedback. Similarly, it seems that our simulations are missing one or more important features that work in concert with Type II supernova feedback to regulate a Milky Way-like galaxy. While there may be evidence for weak coupling between gas inflow and star formation, as discussed in Section \ref{sec:self-reg}, we believe that our simulations are likely not capturing physical features or effects that would make this coupling more obvious and/or tighter. 

The biggest evidence for this is the large feedback-driven bubbles that are present in all of our simulation variants with stellar feedback (i.e., excluding the \textsc{CoolFlow} variant). Though the ramp in feedback efficiency, described in Section \ref{sec:FB-ramp}, delays the onset of these bubbles, we don't expect further tweaks to the ramp (such as modifications to its start and end points in both time and efficiency values) to significantly affect the presence of these large bubbles. This is because the dramatic cavities seen in at late times in Figures \ref{fig:fid_ev} and \ref{fig:edge_comp} are due to successive bubbles being blown into the same volume. Earlier feedback winds clear away material, making it easier for later winds to travel faster and farther, and to further remove material from the bubble's cavity. 

These bubbles travel well beyond the virial radius by the end of the simulations at $t=4$~Gyr, leaving cavities of high entropy gas that will not cool within a Hubble time. The density of the ambient medium continues to decrease beyond the virial radius (Figure \ref{fig:ics}), providing nothing for the feedback-driven bubbles to meaningfully collide with and halt against. Additionally, the cooling time at large radii is high enough that material cannot ``backfill'' the bubble cavities. The behavior of these outflows suggest that we are missing either material that the winds would stall against, a mechanism which would fill the cavities in with low entropy gas, or both.

The idealized nature of our simulations may be working against us in this regard. \citet[]{fielding20-FirstResultsSMAUG} compared idealized and cosmological simulations of Milky Way-like galaxies and found that the outer CGM structure of the latter ($\gtrsim0.5 r_{200}$) was highly impacted by cosmological effects such as nonspherical gas accretion and the presence of satellites. No attempt to model cosmological accretion was included in our simulations, but these could address the large wind-driven bubbles in our simulations. This is especially true considering that the FOGGIE simulations \citep{peeples19-FiguringOutGas} use the same feedback algorithm and a slightly \textit{higher} feedback efficiency but do not see the same long-term disruption due to outflows. Appropriately modeling inflow in an idealized simulation is a challenge, however. Typically, cosmological accretion in idealized simulations is treated as being spherically symmetric, as in \citet[]{fielding17-ImpactStarFormation} (albeit with added density fluctuations), but as stated above, \citet[]{fielding20-FirstResultsSMAUG} emphasized that this spherical treatment is not sufficient. That work highlighted the strength of idealized simulations for studying the inner CGM ($\lesssim0.5r_{200}$), but our work indicates that for self-regulation the outer CGM and perhaps the nearby IGM exert an important influence.

Our initial CGM is also smooth and spherically symmetric. A more realistic CGM would be properly multiphase, with fluctuations in density, temperature, and velocity (both radial and tangentially). We initially expected to see such fluctuations develop over the course of the simulation as a result of outflows, but including these from the start would likely have a profound effect. A nonuniform CGM will have a distribution of cooling times at a given radius, better mixing, and may be better able to disrupt early outflows. In the simulations presented in this work, the earliest outflows are able to expand through the CGM nearly uniformly.

Initial density perturbations are the easiest way to disrupt this spherical symmetry, but require a choice of power spectrum. A more natural way of introducing perturbations at large radii may occur from adjusting our CGM beyond $r_{200}$. Our current treatment for this ``outer'' gas was motivated by practical considerations rather than observations. Notably, our entropy initial profile increases monotonically with radius. More realistic halos are likely to have low-entropy gas beyond $r_{200}$: gas that has either not yet passed through a cosmological accretion shock, or has passed through the shock with enough density for its current entropy to be less than the mean at $r_{200}$. The relaxation of the simulation's initial state could push this gas to become Rayleigh-Taylor unstable, and either over time or with the assistance of instabilities seeded in the initial conditions, could grow dense clumps of gas that may affect the outflows and overall precipitation within our simulations.

Cosmological inflow would further assist in seeding and/or maintaining CGM fluctuations, naturally creating them as infalling material combines with the galaxy. This is true both for filamentary inflow as well as the presence of and mergers with small satellites and dwarf galaxies.

The failure of our simulations to ``close the feedback loop'' is not entirely novel. \citet[]{prasad20-EnvironmentalDependenceSelfregulating} simulated an elliptical galaxy of similar mass ($2\times 10^{12}~\Msun$), dominated by AGN feedback, and also saw a highly disrupted CGM emerge. Generally, lower mass galaxies have a weaker gravitational potential and lower CGM pressure. Outflows can more easily escape the halo. This reinforces our earlier observation that some physical mechanism at the edge of the halo---cosmological inflows or simply more generic density perturbations, to name two candidates---appears necessary for lower mass halos to maintain their CGM structure.

Though we suspect the lack of cosmological effects is the biggest missing piece from our simulations, there are other physical processes to consider. Our current simulations use Type II supernovae as their only feedback source. Though our simulations do not currently suffer from a lack of outflows, efforts to counteract their current behavior may highlight the need to include other forms of feedback, such as Type Ia supernovae and AGN. We adopt a fairly straightforward prescription for rotation in the CGM and assume its angular momentum is aligned with that of the disk, but the angular momentum of the CGM is likely quite complicated \citep[][]{cadiou21-GravitationalTorquesDominate}. This could facilitate better mixing between outflows and the ambient medium, disrupting the structures that develop in our simulations. 

Finally, our simulations omit two important plasma components: magnetic fields and cosmic rays. These two influences would significantly alter the behavior of outflows. Magnetic fields can slow outflows and raise the density of the inner CGM, though they also hinder metal mixing \citep[][]{vandevoort21-EffectMagneticFields}. Cosmic rays provide a form of non-thermal pressure support, allowing cold gas to occupy more CGM volume \citep[][]{ji20-PropertiesCircumgalacticMedium}. Cosmic rays also lead to larger, lower density cold clouds, and keep cold gas in the CGM for longer \citep[][]{butsky20-ImpactCosmicRays}. All of these changes have implications for the galaxy's accretion of CGM gas.

Disruption is not a universal feature of CGM-focused idealized simulations. \citet[]{fielding17-ImpactStarFormation} and \citet[]{li20-HowSupernovaeImpact} do not include a cold gas disk nor explicitly model star formation. Instead, they tie outflows to the amount of inflow through an inner boundary. In both cases, outflows do not cause a large-scale disruption of the CGM. \citet[]{li20-HowSupernovaeImpact} see a clear net balance between gas inflow and outflow in the latter half of their simulation, and both works observe clear cold gas condensation near the galaxy ($r \lesssim 100$~kpc). Yet tying outflows directly to inflows, rather than depending on the intermediary process of star formation, may under-predict the strength of stellar feedback because it neglects star formation due to gas already present in the disk. It also inputs feedback energy into a region that is spatially removed from star formation, where gas is more likely to be lower density.

\section{Conclusions}\label{sec:conclusions}

We have run a suite of isolated, idealized Milky Way-like galaxy simulations in order to examine the ability of galaxies to self-regulate their star formation. They were specifically designed to explore the precipitation theory of self-regulation \citep{voit15-PrecipitationRegulatedStarFormation}. The circumgalactic medium (CGM) in our galaxies was initialized in hydrostatic equilibrium with entropy profiles set by expectations from precipitation \citep{voit19-AmbientColumnDensities}. The CGM was also given an initial azimuthal rotation scaled off the estimates of \citet[]{hodges-kluck16-RotationHotGas}. We explored variations in the entropy profile through the precipitation limit parameter $\tau$ in Equation \ref{eq:entropy} as well as variations in the rotation profile of the CGM (Equation \ref{eq:rot}).

Ours are the first idealized simulations to include both the CGM and explicit star formation in a rotating disk. Previous idealized CGM simulations \citep{fielding17-ImpactStarFormation, li20-HowSupernovaeImpact} tied outflow directly to inflow through an inner boundary. Explicitly modeling star formation brings additional challenges, as the cold gas disk typically used as an initial condition in isolated galaxy models leads to large initial burst of star formation. Outflows from the resulting feedback quickly disrupt the CGM. This has historically not been a challenge for isolated galaxy simulations \citep[e.g.,][]{kim16-AGORAHighresolutionGalaxy, benincasa16-AnatomyStarformingGalaxy} because they have had an essentially non-existent CGM. To prevent the initial star formation burst from disrupting our CGM before its gas can accrete onto the disk, we implement a ramp in the stellar feedback's efficiency parameter (Section \ref{sec:FB-ramp}). This ramp minimizes the impact of feedback for the first 1~Gyr. As a result, the impacts of our CGM variations become apparent.

Our work highlights that including explicit star formation in idealized simulations is crucial to understanding the galaxy-CGM connection. Alternative solutions, such as tying outflow rates directly to gas accretion rates, obfuscate important steps connecting accretion and feedback.
We ultimately fail to produce isolated galaxies that are able to self-regulate their star formation, but this failure is illuminating in several ways. The primary results of our simulations are as follows:

\begin{itemize}
    
    \item Idealized galaxy simulations are highly sensitive to their initial conditions. Our simulation setup is not unusual among isolated disk galaxy simulations, but complications arise when including the CGM. Chief among these is a large initial burst of star formation that can disrupt the CGM.
    
    \item Even after mitigating the initial star formation burst, our simulations contain outflows that are very disruptive to the CGM. Disruptive feedback seems to be a common feature of isolated galaxy simulations at this halo mass \citep{prasad20-EnvironmentalDependenceSelfregulating}. This indicates that current idealized simulations are missing important features that would constrain, disrupt, or backfill these outflows, such as cosmological inflow. 
    
    \item Our galaxies continue to accrete gas along the midplane of the disk, despite the disruption of the CGM by outflows. Though this accretion channel contains gas with the lowest $\tctff$ ratio, it is still higher than expected for a precipitation-regulated system. This accretion is able to maintain low SFRs of $\sim0.1\ \Msun$/yr.
    
    \item Rotation in the CGM impacts the ability of gas to accrete onto the disk. This accretion also varies with the rotation profile. Rotation is therefore an important component of idealized CGM studies, and better understanding angular momentum in the CGM is an important prerequisite to informative modeling.
\end{itemize}

Understanding the balance between accretion rate and star formation first requires an understanding of the time scales over which we expect these processes to balance. While our work is not able to answer this question, we find it important for the community to consider as studies of the CGM's impact on star formation continue.
    
Future work will incorporate some of the physical features we have identified as potentially important for mitigating the dominance of outflows in our idealized simulations. We also wish to investigate the importance of magnetic fields and cosmic rays, particularly since the latter can dramatically affect the CGM's cold gas fraction \citep{ji20-PropertiesCircumgalacticMedium,butsky18-RoleCosmicrayTransport}.

\acknowledgements{
CK was supported in part by a Department of Energy Computational Science Graduate Fellowship under Award Number DE-FG02-97ER25308.
CK and BWO acknowledge support from NSF grant \#1908109, and BWO acknowledges further support from NASA ATP grants NNX15AP39G and 80NSSC18K1105. GMV and BWO acknowledge support from NSF grant \#2106575.
This work used the Extreme Science and Engineering Discovery Environment (XSEDE) under allocation TG-AST090040, as well as the resources of the Michigan State University High Performance Computing Center (operated by the Institute for Cyber-Enabled Research).}

\software{NumPy \citep{harris20-ArrayProgrammingNumPy}, SciPy \citep{virtanen20-SciPyFundamentalAlgorithms}, Matplotlib \citep{hunter07-Matplotlib2DGraphics}, Pandas \citep{mckinney10-DataStructuresStatistical}, yt \citep{turk11-YtMulticodeAnalysis}, and GNU parallel \citep{tange20-GNUParallel20200822}.}

\bibliography{CGM}{}
\bibliographystyle{aasjournal}
\end{document}